\documentclass[11pt,english]{article}
\usepackage[T1]{fontenc}
\usepackage[latin9]{inputenc}
\usepackage{geometry}
\usepackage[active]{srcltx}
\usepackage{babel}
\usepackage{verbatim}
\usepackage{float}
\usepackage{textcomp}
\usepackage{amsmath}
\usepackage{amsthm}
\usepackage{amssymb}
\usepackage{graphicx}
\usepackage{setspace}
\usepackage[authoryear]{natbib}
\usepackage[unicode=true]
 {hyperref}

\makeatletter

\providecommand{\tabularnewline}{\\}
\floatstyle{ruled}
\newfloat{algorithm}{tbp}{loa}
\providecommand{\algorithmname}{Algorithm}
\floatname{algorithm}{\protect\algorithmname}

\theoremstyle{definition}
\newtheorem{problem}{\protect\problemname}
\theoremstyle{plain}
\newtheorem{thm}{\protect\theoremname}
\theoremstyle{definition}
\newtheorem*{defn*}{\protect\definitionname}
\theoremstyle{definition}
\newtheorem*{example*}{\protect\examplename}
\theoremstyle{plain}
\newtheorem{lem}{\protect\lemmaname}
\theoremstyle{plain}
\newtheorem{assumption}{\protect\assumptionname}
\theoremstyle{plain}
\newtheorem{cor}{\protect\corollaryname}
\theoremstyle{definition}
 \newtheorem{example}{\protect\examplename}
\theoremstyle{remark}
\newtheorem{rem}{\protect\remarkname}
\theoremstyle{remark}
\newtheorem*{claim*}{\protect\claimname}

\usepackage{fullpage}
\usepackage{verbatim}
\usepackage{diagbox}

\allowdisplaybreaks
\usepackage{hyperref}
\hypersetup{
     colorlinks   = true,
     citecolor    = blue
}
\date{}

\makeatother

\providecommand{\assumptionname}{Assumption}
\providecommand{\claimname}{Claim}
\providecommand{\corollaryname}{Corollary}
\providecommand{\definitionname}{Definition}
\providecommand{\examplename}{Example}
\providecommand{\lemmaname}{Lemma}
\providecommand{\problemname}{Problem}
\providecommand{\remarkname}{Remark}
\providecommand{\theoremname}{Theorem}

\begin{document}
\global\long\def\Acal{\mathcal{A}}%

\global\long\def\Bcal{\mathcal{B}}%

\global\long\def\Ccal{\mathcal{C}}%

\global\long\def\bC{\bar{C}}%

\global\long\def\Dcal{\mathcal{D}}%

\global\long\def\te{\tilde{e}}%

\global\long\def\tF{\tilde{F}}%

\global\long\def\bF{\bar{F}}%

\global\long\def\hF{\hat{F}}%

\global\long\def\hf{\hat{f}}%

\global\long\def\bG{\bar{G}}%

\global\long\def\hG{\hat{G}}%

\global\long\def\tG{\tilde{G}}%

\global\long\def\Hcal{\mathcal{H}}%

\global\long\def\hL{\hat{L}}%

\global\long\def\Lcal{\mathcal{L}}%

\global\long\def\Kcal{\mathcal{K}}%

\global\long\def\Mcal{\mathcal{M}}%

\global\long\def\tM{\tilde{M}}%

\global\long\def\bM{\bar{M}}%

\global\long\def\hM{\hat{M}}%

\global\long\def\hr{\hat{r}}%

\global\long\def\tR{\tilde{R}}%

\global\long\def\bR{\bar{R}}%

\global\long\def\RR{\mathbb{R}}%

\global\long\def\SS{\mathbb{S}}%

\global\long\def\Tcal{\mathcal{T}}%

\global\long\def\hU{\hat{U}}%

\global\long\def\tu{\tilde{u}}%

\global\long\def\hV{\hat{V}}%

\global\long\def\bW{\bar{W}}%

\global\long\def\tW{\tilde{W}}%

\global\long\def\tw{\tilde{w}}%

\global\long\def\ty{\tilde{y}}%

\global\long\def\tZ{\tilde{Z}}%

\global\long\def\agmin{\arg\min}%

\global\long\def\hbeta{\hat{\beta}}%

\global\long\def\tbeta{\tilde{\beta}}%

\global\long\def\bbeta{\bar{\beta}}%

\global\long\def\teps{\tilde{\varepsilon}}%

\global\long\def\heps{\hat{\varepsilon}}%

\global\long\def\hpi{\hat{\pi}}%

\global\long\def\hxi{\hat{\xi}}%

\global\long\def\txi{\tilde{\xi}}%

\global\long\def\tDelta{\tilde{\Delta}}%

\global\long\def\tdelta{\tilde{\delta}}%

\global\long\def\hdelta{\hat{\delta}}%

\global\long\def\hgamma{\hat{\gamma}}%

\global\long\def\hGamma{\hat{\Gamma}}%

\global\long\def\tGamma{\tilde{\Gamma}}%

\global\long\def\bGamma{\bar{\Gamma}}%

\global\long\def\hSigma{\hat{\Sigma}}%

\global\long\def\bSigma{\bar{\Sigma}}%

\global\long\def\tSigma{\tilde{\Sigma}}%

\global\long\def\hTheta{\hat{\Theta}}%

\global\long\def\dTheta{\dot{\Theta}}%

\global\long\def\bTheta{\bar{\Theta}}%

\global\long\def\cTheta{\check{\Theta}}%

\global\long\def\rTheta{\mathring{\Theta}}%

\global\long\def\hPsi{\hat{\Psi}}%

\global\long\def\tXi{\tilde{\Xi}}%

\global\long\def\lnorm{\left\Vert \right\Vert }%

\global\long\def\abs{\left|\right|}%

\global\long\def\htheta{\hat{\theta}}%

\global\long\def\dv{\dot{v}}%

\global\long\def\rank{{\rm rank}\,}%

\global\long\def\trace{{\rm trace}}%

\global\long\def\boldone{\mathbf{1}}%

\global\long\def\hmu{\hat{\mu}}%

\global\long\def\hA{\hat{A}}%

\global\long\def\hq{\hat{q}}%

\global\long\def\Fcal{\mathcal{F}}%

\global\long\def\tmu{\tilde{\mu}}%

\global\long\def\tX{\tilde{X}}%

\global\long\def\bX{\bar{X}}%

\global\long\def\tx{\tilde{x}}%

\global\long\def\htau{\hat{\tau}}%

\global\long\def\eig{{\rm eig}}%

\global\long\def\tq{\tilde{q}}%

\global\long\def\ttau{\tilde{\tau}}%

\global\long\def\oneb{\mathbf{1}}%

\global\long\def\ZZ{\mathbb{Z}}%

\global\long\def\QQ{\mathbb{Q}}%

\title{Learning non-smooth models: instrumental variable quantile regressions
and related problems\thanks{I would like to thank Stéphane Bonhomme, Victor Chernozhukov, Christian
Hansen, David Kaplan, Whitney Newey, Lawrence Schmidt and Kaspar Wüthrich
for their comments and discussions.}}
\author{Yinchu Zhu\thanks{Email: yzhu6@uoregon.edu}\\
\textit{University of Oregon}}
\maketitle
\begin{abstract}
This paper proposes computationally efficient methods that can be
used for instrumental variable quantile regressions (IVQR) and related
methods with statistical guarantees. This is much needed when we investigate
heterogenous treatment effects since interactions between the endogenous
treatment and control variables lead to an increased number of endogenous
covariates. We prove that the GMM formulation of IVQR is NP-hard and
finding an approximate solution is also NP-hard. Hence, solving the
problem from a purely computational perspective seems unlikely. Instead,
we aim to obtain an estimate that has good statistical properties
and is not necessarily the global solution of any optimization problem. 

The proposal consists of employing $k$-step correction on an initial
estimate. The initial estimate exploits the latest advances in mixed
integer linear programming and can be computed within seconds. One
theoretical contribution is that such initial estimators and Jacobian
of the moment condition used in the $k$-step correction need not
be even consistent and merely $k=4\log n$ fast iterations are needed
to obtain an efficient estimator. The overall proposal scales well
to handle extremely large sample sizes because lack of consistency
requirement allows one to use a very small subsample to obtain the
initial estimate and the $k$-step iterations on the full sample can
be implemented efficiently. Another contribution that is of independent
interest is to propose a tuning-free estimation for the Jacobian matrix,
whose definition nvolves conditional densities. This Jacobian estimator
generalizes bootstrap quantile standard errors and can be efficiently
computed via closed-end solutions. We evaluate the performance of
the proposal in simulations and an empirical example on the heterogeneous
treatment effect of Job Training Partnership Act.%

\thispagestyle{empty}
\end{abstract}
\pagebreak{}

\setcounter{page}{1}

\section{\label{sec:Introduction}Introduction}

The linear instrumental variables quantile model (IVQR) formulated
by \citet{chernozhukov2005iv,chernozhukov2006instrumental,Chernozhukov2008}
has found wide applications in economics. The basic moment condition
can be written as follows
\begin{equation}
Y_{i}=X_{i}'\beta_{*}+\varepsilon_{i}\qquad{\rm and}\qquad P(\varepsilon_{i}\leq0\mid Z_{i})=\tau,\label{eq: linear IV quantile model}
\end{equation}
where $\tau\in(0,1)$ is the quantile of interest, $Y_{i}\in\mathbb{R}$,
$X_{i}\in\mathbb{R}^{p}$ and $Z_{i}\in\mathbb{R}^{L}$ are i.i.d
observed variables and $\beta_{*}\in\mathbb{R}^{p}$ is the unknown
model parameter. Assume that $p$ and $L$ are fixed with $L\geq p$.
The typical setup is that only one (or few) component of $X_{i}$
is endogenous and other components of $X_{i}$ are contained in $Z_{i}$.
In the policy evaluation setting, the variable denoting the status
of treatment is usually considered endogenous. If this variable only
enters the regression equation as one endogenous regressor, then we
can apply the existing methods (e.g., the popular method by \citet{chernozhukov2006instrumental})
for estimating the treatment effect. When multiple endogenous regressors
enter the regression equation, it imposes enormous (or even prohibitive)
computational challenges to common estimation strategies, which typically
involve solving nonconvex and non-smooth optimization problems. %

In empirical studies that investigate the causal effects of an endogenous
treatment, multiple endogenous regressors arise naturally. For example,
the interactions between the treatment variable and other variables
are often included in the regression to study the heterogeneity of
the treatment effects. In the IVQR setting, this leads to multiple
endogenous variables in the regression equation. Consider the randomized
training experiment conducted under the Job Training Partnership Act
(JTPA). JTPA training services are randomly offered to people, who
can then choose whether to participate in the program. One key policy
question is whether this program has an effect on earnings. Of course,
the baseline question is whether the program has a positive effect
overall. In addition, one might ask questions such as whether the
effect of the program differs by participants' race, age, etc. These
questions can be answered in the regression setting by learning the
coefficients for interactions between the treatment status and other
variables denoting race, age, etc. %

We show that the efficient computations for estimating IVQR from a
purely algorithmic perspective might not exist. In particular, we
prove that optimization of the usual GMM or similar formulation of
the IVQR problem is NP (non-deterministic polynomial-time) hard. To
the best of our knowledge, this is the first formal statement on the
NP-hardness of the IVQR problem. Due to this result, it is unlikely
to find a computationally efficient algorithm that minimizes the GMM
criterion function. Notice that this is different from some existing
methods that cast the GMM problem as NP-hard problems, e.g., mixed
integer program or quadratic program with complementarity constraints
\citep{burer2012milp,chen2017exact}. These papers only imply that
these NP-hard problems are more difficult than IVQR GMM-estimation;
in other words, algorithms that solve certain NP-hard problems can
be used for IVQR estimation. However, the question of whether there
exist simple algorithms for IVQR GMM-estimation is still open. In
this paper, we provide a negative answer by showing that IVQR estimation
is at least as difficult as the set partition problem, which is known
to be NP-hard since \citet{karp1972reducibility}. Moreover, we show
that it is also very difficult to find an approximate solution in
that obtaining a solution within a constant factor from the global
solution is also NP-hard. Therefore, there is inherent complexity
that is unlikely to be fully handled by convex relaxation, smoothing
or other pure computational remedies for handling non-convex and non-smooth
optimization problems. %

The NP-hardness suggests that serious efforts are needed to address
the computational challenge, but it does not necessarily mean that
the model has to be shunned by practitioners. In fact, some of the
most popular methods, such as decision trees and $k$-means clustering
\citep{laurent1976constructing,MAHAJAN201213}, are NP-hard. One key
observation is that NP-hardness is a statement on the difficulty of
the worst-case situation, but the data that arise in practice might
not be these ``bad'' situations. Although optimization algorithms
derived from a pure computational point of view are not guaranteed
to exploit ``probabilistic'' properties of the data, there are many
inspiring precedents in machine learning that exploit statistical
properties to overcome computational difficulties. For example, finding
a hyperplane to classify binary targets is a common approach. Whereas
using the misclassification rate as the loss function leads to an
NP-hard optimization problem, one can use a convex loss function (e.g.,
hinge loss or logistic loss) and still obtain nice statistical properties,
see \citep{bishop2006pattern,shalev2014understanding}. 

Here, we propose to apply the same principle to general non-smooth
models. Instead of focusing only on clever computational tricks for
the usual GMM formulation, we exploit the statistical aspect of the
model and provide an alternative estimation and inference strategy
that admits fast computation algorithms. Although estimates from any
feasible optimization programs might be of questionable quality, our
plan is to use a computationally cheap method and turn a dubious estimate
into one with nice statistical properties. Of course, this final estimate
might not be the global solution of any optimization problem but good
statistical properties would often suffice for estimation and inference
purposes. 

Our proposal is a $k$-step framework that exploits the identification
strength of a general (possibly nonsmooth) GMM model.\footnote{In a sense, many methods for high-dimensional sparse estimation can
be also viewed as exploiting identification in order to avoid computational
difficulties. Due to the NP-hardness of computing ``$\ell_{0}$-regularized''
estimators \citep[e.g.,][]{chen2014complexity}, many $\ell_{1}$-regularized
estimators have been considered and, under strong identification (in
the form of sparse eigenvalue conditions or similar conditions), have
been shown to achieve optimal statistical properties \citep[see e.g.,][]{candes2007dantzig,RaskuttiWainwrightYu2011,cai2017confidence}.} This framework starts with an initial estimate and updates it iteratively,
where each iteration involves only matrix multiplications and no optimization
at all. We show that when the identification is strong, only few iterations
are needed to obtain an estimator that is asymptotically equivalent
to the GMM estimator.

One key finding is that the initial estimator does not need to be
consistent. This is a new result as far as we know. We show that there
is a neighborhood of the true parameter value whose area is determined
by the identification strength and any point in this neighborhood
can be used as an initial estimate; when the identification is strong
in the usual sense for GMM models (i.e., Jacobian of the population
moment has well-behaved singular values), this neighborhood is fixed
and its area does not shrink to zero with sample size $n\rightarrow\infty$.
Allowing for inconsistent initial estimates provides valuable convenience
in practice. When the initial estimates are computed via complicated
optimization problems, computationally feasible solutions might not
be consistent since they are often not global solutions even under
strong identification. Moreover, even if initial estimates are easily
obtained via efficient algorithms, these algorithms might not be feasible
when the sample size becomes extremely large. For large datasets that
arise in modern research, even common convex problems often encounter
scalability issues, see e.g., \citep{yang2013quantile,lin2017distributed}.
Since the initial estimate does not need to be even consistent, one
can simply compute it using a very small fraction of the entire sample.
Efficient algorithms are available for $k$-step iterations as they
only require matrix multiplications. This leads to a fast procedure
for handling massive datasets.

We also provide methods and theory for two important aspects of implementing
the $k$-step framework. First, we provide a tuning-free estimate
for the Jacobian matrix of the moment condition. This estimate is
needed for the $k$-step correction and can be tricky to obtain in
practice. The formula defining the Jacobian typically contains the
(conditional) density of a variable and one typical estimation method
is to use kernel with a bandwidth choice. However, the bandwidth might
be quite difficult to choose in practice. We propose to generalize
a fast bootstrap scheme for density estimation. The basic idea is
as follows. Consider the estimation of the density of a variable at
a point. Since the asymptotic variance of the bootstrapped sample
quantile of this variable is directly related to the density, bootstrapping
sample quantile can be used to obtain a consistent density estimate.
We generalize this approach to estimation of Jacobian of general nonsmooth
GMM models. Notice that the Jacobian matrix can be estimated entry
by entry. In the case of IVQR, this has a closed-end solution with
computation burden similar to bootstrapping sample quantiles; in our
experiments, with a sample size $n=5000$, computing 2000 bootstrap
samples takes 0.5 second. Since the entire procedure can be used for
general GMM settings and does not require choices such as bandwidth,
we believe it is of independent interest. For example, one can use
it as a tuning-free density estimator, which can be used to compute
the optimal bandwidth for a second-stage estimation. We establish
the consistency for this estimate. %

Second, we present a formulation that yields initial estimators for
IV quantile models and related problems via mixed integer linear programs
(MILP). Although MILP is also an NP-hard problem, it is one of the
most well-studied and well-understood hard problems and so much progress
has been made in recent years that many believe large-scale problems
are now feasible, see e.g., \citep{bertsimas2005optimization,bixby2007progress,junger200950,linderoth2010milp}.
As pointed out by \citet{bertsimas2016best}, the speed of finding
global solutions for mixed integer optimization improved approximately
450 billion times between 1994 and 2015. In our experiments, we deliver
good estimates for coefficients of 20 endogenous variables within
5 seconds. The high-dimensional version can handle regression equations
with 500 endogenous variables within minutes. Since we do not need
the initial estimate to be consistent, one can terminate MILP optimization
before a global solution is found. We also provide an early-termination
rule and establish its theoretical validity. The MILP constructions
are not unique to low-dimensional IV quantile regressions. We outline
how MILP can be used for related problems, including high-dimensional
IV quantile regressions, censored regressions and censored IV quantile
regressions. %

\subsection{\label{subsec: literature review}Related work}

This paper is related to the literature of analyzing computational
complexity of statistical learning methods. One strand of this literature
is based on the NP-completeness theory dating back to at least \citep{cook1971complexity,karp1972reducibility,levin1973universal}.
The classical textbook by \citet{johnson1979computers} documented
hundreds of NP-complete problems. The efficient solution of any of
these problems still remains elusive today and is now widely considered
impossible. Since NP-hard problems are at least as hard as these difficult
problems, finding out whether a statistical procedure is computationally
NP-hard has been an important task for analyzing the implementability,
see e.g., \citet{laurent1976constructing,berthet2013optimal,chen2014complexity,chen2017strong}.
Another recent active area of research is to find efficient approximation
algorithms for NP-hard problems and characterize theoretical possibilities
of such approximations, see \citep{arora2009computational,williamson2011design}
for excellent textbook treatments and surveys.

An early version of this paper was inspired by the fascinating literature
of applying mixed integer programming to statistical learning. Recent
progress has drastically improved the speed of mixed integer optimizations,
which are now considered a feasible tool for some high-dimensional
problems. Most of the advancement concerns high-dimensional linear
models; see \citet{bertsimas2014least,liu2016global,bertsimas2016best,mazumder2017discrete}.
The main argument for considering these nonconvex algorithms is that
they, compared to convex regularized methods, enjoy more desirable
statistical properties. \citet{zubizarreta2012using} proposed using
mixed integer programming for matching estimators in causal inference.

Our work contributes to the fast growing literature of IV quantile
regression. The IV quantile regression extends the advantage of quantile
regression (\citet{Koenker1978}) to the settings with endogenous
regressors. The conceptual framework and identification of the IV
quantile models has been studied by \citet{abadie2002instrumental},
\citet{chernozhukov2005iv} and \citet{imbens2009identification};
see \citet{wuthrich2014comparison,wuthrich2019closed}, \citet{melly2016local}
and \citet{chernozhukov2017instrumental} for more discussions. Semiparametric
and nonparametric specifications have been studied in \citep{horowitz2007nonparametric,chernozhukov2007instrumental,chen2009efficient,chen2012estimation,gagliardini2012nonparametric}.
The GMM estimation approach applies the classical GMM method for the
moment condition in (\ref{eq: linear IV quantile model}). The computational
burden of minimizing a nonconvex and non-smooth objective function
is known to be challenging for larger dimensional models. The quasi-Bayesian
approach of \citet{chernozhukov2003mcmc} has been suggested, but
could be difficult to tune it to sufficiently explore the entire parameter
space. In an interesting paper, \citet{chen2017exact} proposed formulating
the original GMM problem as a mixed integer quadratic program (MIQP).
Smoothing the GMM objective function has also been considered by \citet{kaplan2017smoothed}
and \citet{deCastroGalvaoKaplanLiu2018}. The so-called inverse quantile
regression by \citet{chernozhukov2006instrumental,Chernozhukov2008}
takes a different route and reduces the dimension of the space over
which the optimization is needed. \citet{lee2007endogeneity} considers
a control function approach but deviates from the model (\ref{eq: linear IV quantile model}).
\citet{1kaido2018} propose a decomposition of the IVQR problem into
a set of convex sub-problems. \citet{pouliot2019} develops mixed
integer programs that avoid nonparametric density estimation and allow
for weak identification. 

Our work is also related to the $k$-step estimator in the econometrics
and statistics literature. The classical references include \citet{robinson1988stochastic}
and \citet{andrews2002equivalence}. The main difference in assumption
is that our results do not assume that the sample version of the moment
condition is differentiable. We provide a general theory in this setting,
which might be of independent interest. Moreover, we show that a consistent
starting point is not necessary. 

We will use $E_{n}$ to denote the sample average $n^{-1}\sum_{i=1}^{n}$.
The $\ell_{q}$-norm of a vector will be denoted by $\|\cdot\|_{q}$
for $q\geq1$; $\|\cdot\|_{\infty}$ denotes the maximum absolute
value of a vector, i.e., the $\ell_{\infty}$-norm. Hence, $\|\cdot\|_{2}$
denotes the Euclidean norm. We use $\|\cdot\|$ to denote the spectral
norm of a matrix. The indicator function is denoted by $\mathbf{1}\{\}$.
For any positive integer $r$, we use $\mathbf{1}_{r}$ to denote
the $r$-dimensional vector of ones and $I_{r}$ to denote the $r\times r$
identity matrix. We use $\lambda_{\max}(\cdot)$ and $\lambda_{\min}(\cdot)$
to denote the maximal and the minimal eigenvalues of symmetric matrices.
We use $\log$ to denote the natural logarithm. The rest of the paper
is organized as follows. Section \ref{sec: k step} provides a general
theory of $k$-step correction for non-smooth problems and outlines
the details of implementation for IVQR. Section \ref{sec: IVQR} presents
the MILP formulation of IVQR and related problems; we also derive
theroetical results for early termination of the algorithm without
needing to find a global solution. Section \ref{sec: Jacobian estimate}
presnets a new tuning-free methodology for Jacobian estimation, which
is of independent interest. Monte Carlo simulations are presented
in Section \ref{sec:Monte-Carlo-simulations}. Section \ref{sec: Empirical-Illustration}
considers the JTPA example. The proofs of theoretical results are
in the appendix.

\section{\label{sec: k step}$k$-step correction for non-smooth problems}

\subsection{NP-hardness of IVQR via GMM}

We start by considering the computational complexity of the GMM formulation
of IVQR. The moment condition for (\ref{eq: linear IV quantile model})
is 
\[
EZ_{i}\left(\oneb\{Y_{i}-X_{i}'\beta\}-\tau\right)=0.
\]

Of course, we can replace $Z_{i}$ with transformations of $Z_{i}$;
here, we assume that $Z_{i}$ is already the desired transformation
of the instruments. Then the GMM estimator is 
\[
\hbeta_{GMM}=\underset{\beta\in\RR^{p}}{\arg\min}\left\Vert \hat{\Omega}^{1/2}\sum_{i=1}^{n}Z_{i}\left(\oneb\{Y_{i}-X_{i}'\beta\}-\tau\right)\right\Vert _{2},
\]
where $\hat{\Omega}$ is a weighting matrix, either estimated from
the data or pre-determined. For simplicity, we use the identity $\hat{\Omega}=I_{r}$.
Then the computation can be summarized as follows.
\begin{problem}
\label{prob: original IVQR}Given a number $\tau\in(0,1)$ and data
$\{(x_{i},z_{i},y_{i})\}_{i=1}^{n}$ with $x_{i}\in\RR^{p}$, $z_{i}\in\RR^{L}$
and $y_{i}\in\RR$, we would like to solve 
\[
\min_{\beta\in\RR^{p}}\ \left\Vert \sum_{i=1}^{n}z_{i}\left(\oneb\{y_{i}-x_{i}'\beta\leq0\}-\tau\right)\right\Vert _{2}.
\]
\end{problem}
This seemingly routine GMM estimation turns out to be NP-hard. The
characterization of NP-hardness is an important assessment on the
computational complexity of a problem. Loosely speaking, the class
of NP is the set of problems for which a potential answer can be verified
efficiently (in polynomial time). Consider the set partition problem:
given $\{a_{i}\}_{i=1}^{m}$ integers, decide whether there exists
a subset $B\subset\{1,....m\}$ such that $\sum_{i\in B}a_{i}=\sum_{i\notin B}a_{i}$.
For any given set $B$, we can easily verify whether $\sum_{i\in B}a_{i}=\sum_{i\notin B}a_{i}$
holds; as a result, the set partition problem is in NP. However, the
question of whether such a set $B$ exists is a notoriously difficult
problem. In fact, the set partition problem belongs to the set of
the hardest NP problems, the so-called NP-completeness class. All
the problems in the NP-completeness class are equivalent to each other
and solving any NP-complete problem would also solve every NP problem.
The NP-completeness class contains many problems that are considered
difficult, such as Boolean satisfiability problem, traveling salesman
problem, set partition problem, vertex cover problem, etc. No algorithms
that can solve an NP-complete problem in polynomial time have been
found and whether such an algorithm exists remains an open question,
one of the most fundamental questions in computer science and modern
mathematics. A detailed treatment on NP-hardness can be found in the
classical textbook \citet{johnson1979computers} or chapter 7 of \citet{sipser2012introduction}.%

\begin{thm}
\label{thm: IVQR NP hard}Problem \ref{prob: original IVQR} is NP-hard.
Moreover, the result also holds even if we change $\|\cdot\|_{2}$-norm
to $\|\cdot\|_{r}$-norm for any $1\leq r\leq\infty$.
\end{thm}
The proof of Theorem \ref{thm: IVQR NP hard} establishes that any
algorithm solving Problem \ref{prob: original IVQR} can also be used
to solve the set partition problem, which is an NP-complete problem.
From the proof, one can see that the difficulty arises from the dimension
$p$ since the problem is already NP-hard when both $n$ and $L$
are equal to $p$. Since there are still no fast algorithms for any
NP-complete problem, we do not expect to magically find a fast algorithm
for Problem \ref{prob: original IVQR} and thereby solve all the NP-complete
problems.%

Since finding the exact solution for NP-hard problems is difficult
(if not impossible), one important question is whether it is possible
to find a good approximation. To be specific, we introduce the following
definition adapted from chapter 3 of \citet{ausiello2012complexity}.
\begin{defn*}
Given a minimization problem and a constant $\rho>1$, an algorithm
is said to be a $\rho$-approximation if for all instances, this algorithm
yields a solution whose objective function value is bounded above
by $\rho$ times the global minimum.
\end{defn*}
In terms of the above criteria, there has been tremendous success
in finding efficient approximation algorithms for many problems, but
satisfactory enough approximation might not be possible. For example,
the $k$-center problem admits a simple $2$-approximation and the
traveling sales problem has a fast $3/2$-approximation; however,
it is NP-hard to find a $\rho$-approximation of the former for any
$\rho<2$ and of the latter for any $\rho<220/219$, see \citet{williamson2011design}.
Unfortunately, there are also problems, such as the set covering problem,
for which an approximation solution is as hard as the exact solution
since a $\rho$-approximation is NP-hard for any $\rho>1$, see \citet{lund1994hardness}.
We now show that the GMM formulation of IVQR belongs to this class
of very difficult optimization problems.
\begin{thm}
\label{thm: NP hard approx IVQR}For any $\rho>1$, it is NP-hard
to find a $\rho$-approximation for Problem \ref{prob: original IVQR}.
Moreover, the result also holds even if we change $\|\cdot\|_{2}$-norm
to $\|\cdot\|_{r}$-norm for any $1\leq r\leq\infty$.
\end{thm}
The proof of Theorem \ref{thm: NP hard approx IVQR} is built on the
idea that any algorithm solving the GMM formulation in Problem \ref{prob: original IVQR}
can be used to solve the so-called minimum unsatisfiability problem
of linear systems. The latter problem is known to have no polynomial-time
$\rho$-approximation for any $\rho>1$. In the machine learning literature,
it is sometimes referred to as the half-space learning problem or
linear perceptron learning problem, which finds the best hyperplane
to classify binary outcomes. A direct consequence of the non-approximability
of this problem is that minimizing the misclassification rate is not
a computationally efficient way of training classifiers, see e.g.,
\citet{ben2001efficient}; for this reason, convex functions, such
as the hinge loss or logistic loss, often serve as the objective function
in these learning problems.

In light of Theorems \ref{thm: IVQR NP hard} and \ref{thm: NP hard approx IVQR},
it appears unrealistic to guarantee adequate econometric properties
by relying exclusively on algorithms that attempt to accurately solve
the GMM formulation. Instead, we focus on designing a procedure that
is based on the statistical properties and is as computationally cheap
as possible. 

\subsection{\label{subsec: general theory k step}$k$-step framework for learning
general GMM models}

We now present our proposal and its theoretical justification. Let
$\{W_{i}\}_{i=1}^{n}$ be i.i.d observations. Let $G(\beta)=Eg(W_{i};\beta)$
be an GMM model, where $g$ is an $\RR^{L}$-valued function that
is possibly non-smooth in $\beta\in\Bcal$. The true parameter value
$\beta_{*}$ is assumed to be uniquely defined by $G(\beta_{*})=0$.
Let $\Gamma_{*}=(\partial G(\beta)/\partial\beta)(\beta_{*})\in\RR^{L\times p}$,
$G_{n}(\beta)=n^{-1}\sum_{i=1}^{n}g(W_{i};\beta)$ and $H_{n}(\beta)=\sqrt{n}(G_{n}(\beta)-G(\beta))$.
\begin{example*}[IVQR]
 In the example of IVQR, we define $g(W_{i};\beta)=Z_{i}(\oneb\{Y_{i}\leq X_{i}'\beta)-\tau)$,
where $W_{i}=(X_{i},Z_{i},Y_{i})$ and $\tau\in(0,1)$ is given.
\end{example*}
We view our proposal as performing two tasks: $\sqrt{n}$-estimation
and inference by asymptotic normality.

\subsubsection{\label{subsec: inconsistency to parametric rate}From an inconsistent
estimator to $\sqrt{n}$-estimation}

Suppose that we have an initial estimator $\bar{\beta}$ for $\beta_{*}$
and an estimator $\hGamma$ for $\Gamma_{*}$. Computing the initial
inputs $\bbeta$ and $\hGamma$ will be addressed in Sections \ref{sec: IVQR}
and \ref{sec: Jacobian estimate}, respectively. Neither $\bbeta$
nor $\hGamma$ is assumed to be consistent. Now consider the following
one-step correction estimator. We define the one-step correction operator
by 
\begin{equation}
\Acal(v,Q)=v-(Q'Q)^{-1}Q'G_{n}(v)\qquad{\rm for}\qquad v\in\RR^{p}\qquad\text{and}\qquad Q\in\RR^{L\times p}.\label{eq: debias operator}
\end{equation}

We also define the sequence $\{\Acal_{k}(v,Q)\}_{k=1}^{\infty}$ recursively
by 
\[
\Acal_{k+1}(v,Q)=\Acal(\Acal_{k}(v,Q),Q),
\]
where $\Acal_{1}(v,Q)=\Acal(v,Q)$. The following result allows us
to compare the estimation errors of $\bbeta$ and its one-step correction.
\begin{lem}
\label{lem: main iter}Let $\Bcal_{0}\subseteq\Bcal$. Suppose that
$\sup_{v\in\Bcal_{0}}\|G(v)-\Gamma_{*}(v-\beta_{*})\|_{2}/\|v-\beta_{*}\|_{2}^{2}\leq c$.
Then for any $\beta\in\Bcal_{0}$, 
\begin{multline*}
\|\Acal(\beta,\hGamma)-\beta_{*}\|_{2}\leq\|(\hGamma'\hGamma)^{-1}\hGamma\|\cdot\|\hGamma-\Gamma_{*}\|\cdot\|\beta-\beta_{*}\|_{2}\\
+\|(\hGamma'\hGamma)^{-1}\hGamma'\|\cdot\left(n^{-1/2}\sup_{v\in\Bcal_{0}}\|H_{n}(v)\|_{2}+c\|\beta-\beta_{*}\|_{2}^{2}\right).
\end{multline*}
\end{lem}
Lemma \ref{lem: main iter} depicts the basic intuition that underlies
the $k$-step estimator. Suppose that 
\[
\rho:=c\|(\hGamma'\hGamma)^{-1}\hGamma'\|\sup_{\beta\in\Bcal_{0}}\|\beta-\beta_{*}\|_{2}+\|(\hGamma'\hGamma)^{-1}\hGamma\|\cdot\|\hGamma-\Gamma_{*}\|<1.
\]

Then Lemma \ref{lem: main iter} implies that 
\[
\|\Acal(\beta,\hGamma)-\beta_{*}\|_{2}\leq\rho\|\beta-\beta_{*}\|_{2}+n^{-1/2}\|(\hGamma'\hGamma)^{-1}\hGamma'\|\sup_{v\in\Bcal_{0}}\|H_{n}(v)\|_{2}.
\]

Define $T_{*}=n^{-1/2}\|(\hGamma'\hGamma)^{-1}\hGamma'\|\sup_{v\in\Bcal_{0}}\|H_{n}(v)\|_{2}/[2(1-\rho)]$.
If $\|\beta-\beta_{*}\|_{2}\geq T_{*}$, then we have $\|\Acal(\beta,\hGamma)-\beta_{*}\|_{2}\leq\bar{\rho}\|\beta-\beta_{*}\|_{2}$,
where $\bar{\rho}=(\rho+1)/2<1$. If $\|\Acal(\beta,\hGamma)-\beta_{*}\|_{2}\geq T_{*}$,
then $\|\Acal_{2}(\beta,\hGamma)-\beta_{*}\|_{2}\leq\bar{\rho}^{2}\|\beta-\beta_{*}\|_{2}$.
As we iterate, the estimation error shrinks exponentially (due to
$\bar{\rho}<1$) until it becomes smaller than $\text{threshold}_{*}$.
Typically, we can establish $T_{*}=O_{P}(n^{-1/2})$. This means that
$k$-step iteration would yield an $\sqrt{n}$-consistent estimator
very quickly.

By induction, we can invoke Lemma \ref{lem: main iter} and obtain
the following result on the rates of convergence for $\Acal_{K}(\beta,\hGamma)$.
For now, we do not consider the randomness yet so the following result
serves as a formalization of the intuition and a finite-sample result
for establishing the final statistical properties.
\begin{thm}
\label{thm: main iter rate} Suppose that $\|\bbeta-\beta_{*}\|_{2}\leq c_{1}$,
$\|\hGamma-\Gamma_{*}\|\leq c_{2}$, $\lambda_{\min}(\hGamma'\hGamma)\geq c_{3}$,
$\sup_{\|v-\beta_{*}\|_{2}\leq c_{1}}\|G(v)-\Gamma_{*}(v-\beta_{*})\|_{2}/\|v-\beta_{*}\|_{2}^{2}\leq c_{4}$
and $\sup_{\|v-\beta_{*}\|_{2}\leq c_{1}}\|H_{n}(v)\|_{2}\leq c_{5}$
such that $c_{5}\leq c_{1}c_{3}^{1/2}(1-\rho_{*})\sqrt{n}$, where
$\rho_{*}=c_{3}^{-1/2}(c_{2}+c_{1}c_{4})$. Then $\rho_{*}<1$ and
for any $K\geq1$, 
\[
\|\Acal_{K}(\bbeta,\hGamma)-\beta_{*}\|_{2}\leq\rho_{*}^{K}c_{1}+n^{-1/2}\frac{c_{3}^{-1/2}c_{5}}{1-\rho_{*}}.
\]
\end{thm}
Theorem \ref{thm: main iter rate} has two important implications.
First, the starting point $(\bbeta,\hGamma)$ does not need to be
a consistent estimator for $(\beta_{*},\Gamma_{*})$. By Theorem \ref{thm: main iter rate}
, whenever we start from a small enough neighborhood (i.e., small
enough $c_{1},c_{2}$), $\|\Acal_{K}(\bbeta,\hGamma)-\beta_{*}\|_{2}$
decays exponentially with $K$ until it reaches the parametric rate
$n^{-1/2}$. The only requirement is that $c_{5}\leq c_{1}c_{3}^{1/2}(1-\rho_{*})\sqrt{n}$.
A sufficient condition is $c_{1}=c_{3}^{1/2}/(2c_{4})$, $c_{2}=c_{3}^{1/2}/2$
and $n\geq16c_{3}^{-2}c_{4}^{2}c_{5}^{2}$. Notice that this requirement
on $c_{1}$ and $c_{2}$ depends on $c_{3}$, which measures the identification
strength. Since $c_{3},c_{4},c_{5}$ are bounded away from zero and
infinity under strong identification, we allow $c_{1}$ and $c_{2}$
to be bounded away from zero and only require $n$ to be large enough
(instead of tending to infinity). %

Second, the parametric rate $n^{-1/2}$ is guaranteed after $O(\log n)$
iterations. Notice that $\rho_{*}<1$. This means that $K\geq(\log n)/(2\log\rho_{*}^{-1})$,
we have that 
\[
\|\Acal_{K}(\bbeta,\hGamma)-\beta_{*}\|_{2}\leq n^{-1/2}\left(c_{1}+\frac{c_{3}^{-1/2}c_{5}}{1-\rho_{*}}\right).
\]
This is computationally quite attractive. Even if the starting point
is not consistent or its rate of convergence can be arbitrarily slow,
we only need a few iterations to obtain a $\sqrt{n}$-consistent estimator.
Since there is no optimization in each iteration, this can be done
extremely fast. In fact, this is the key property we shall exploit
when dealing with massive samples. Now we state the result under commonly
imposed regularity conditions.
\begin{assumption}
\label{assu: main rate}Suppose that the following conditions hold:\\
(1) There exist constants $\kappa_{1},\kappa_{2}>0$ such that $\|G(v)-\Gamma_{*}(v-\beta_{*})\|_{2}\leq\kappa_{1}\|v-\beta_{*}\|_{2}^{2}$
for any $v\in\RR^{p}$ satisfying $\|v-\beta_{*}\|_{2}\leq\kappa_{2}$.
\\
(2) There exists a constant $\kappa_{3}>0$ such that $\lambda_{\min}(\Gamma_{*}'\Gamma_{*})\geq\kappa_{3}$.
\\
(3) $\sup_{\|v-\beta_{*}\|_{2}\leq\kappa_{2}}\|H_{n}(v)\|_{2}=O_{P}(1)$.
\end{assumption}
We have the following result.
\begin{cor}
\label{cor: main rate}Let Assumption \ref{assu: main rate} hold.
Then 
\begin{multline*}
P\left(\sup_{K\geq2\log n}\|\Acal_{K}(\bbeta,\hGamma)-\beta_{*}\|_{2}\leq n^{-1/2}\kappa_{2}+4n^{-1/2}\kappa_{3}^{-1/2}\sup_{\|v-\beta_{*}\|_{2}\leq\kappa_{2}}\|H_{n}(v)\|_{2}\right)\\
\geq1-P\left(\|\hGamma-\Gamma_{*}\|>\frac{\sqrt{\kappa_{3}}}{8}\right)-P\left(\|\bbeta-\beta_{*}\|_{2}>\min\left\{ \frac{\sqrt{\kappa_{3}}}{8\kappa_{1}},\ \kappa_{2}\right\} \right)-o(1).
\end{multline*}
\end{cor}
By Corollary \ref{cor: main rate}, if we have strong identification,
bounded Hessian for $G(\cdot)$ and the empirical process $H_{n}(\cdot)$
is a Donsker class, then after $2\log n$ iterations, we will obtain
a $\sqrt{n}$-consistent estimator as long as $\hGamma$ and $\bbeta$
lie in a small enough but fixed neighborhood of the true parameters
with high probability. Once we obtain a $\sqrt{n}$-consistent estimator
for $\beta_{*}$, we can use it to construct a consistent estimator
for $\Gamma_{*}$; it turns out that the consistency of $\Gamma_{*}$
is needed to obtain asymptotic normality. %

\subsubsection{Further iteration for asymptotic normality}

We now derive the asymptotic normality for the $k$-step estimator.
We also address an important robustness issue. Obviously, the output
of the $k$-step iteration depends on the number of iterations $K$
and the initial estimators $\bbeta$ and $\hGamma$. We now explicitly
express such dependence and address the issue of sensitivity with
respect to $(K,\bbeta,\hGamma)$.
\begin{thm}
\label{thm: inference main}Let Assumption \ref{assu: main rate}
hold. Suppose that $\sup_{\|v\|_{2}\leq C}\|H_{n}(\beta_{*}+n^{-1/2}v)-H_{n}(\beta_{*})\|_{2}=o_{P}(1)$
for any $C>0$. Let $\varepsilon_{n}$ be an arbitrary sequence tending
to zero. Then
\begin{multline}
\sup_{K\geq1+2\log n,\ \|\beta-\beta_{*}\|_{2}\leq A,\ \|\Gamma-\Gamma_{*}\|\leq\varepsilon_{n}}\|\Acal_{K}(\beta,\Gamma)-\beta_{*}+n^{-1/2}(\Gamma_{*}'\Gamma_{*})^{-1}\Gamma_{*}'H_{n}(\beta_{*})\|_{2}\\
\leq O_{P}(\varepsilon_{n}n^{-1/2}+n^{-1})+o_{P}(n^{-1/2}),\label{eq: main inference conclusion}
\end{multline}
where $A=\min\left\{ \sqrt{\kappa_{3}}/(8\kappa_{1}),\ \kappa_{2}\right\} $.
\end{thm}
Theorem \ref{thm: inference main} provides the main tool for inference.
It says that as long as $\hat{\beta}$ and $\hGamma$ are consistent
(easily achieved by first iterating $2\log n$ times from the initial
estimator as shown in Section \ref{subsec: inconsistency to parametric rate}),
we have $\|\Acal_{K}(\hat{\beta},\hGamma)-\beta_{*}+n^{-1/2}(\Gamma_{*}'\Gamma_{*})^{-1}\Gamma_{*}'H_{n}(\beta_{*})\|_{2}=o_{P}(n^{-1/2})$
for any $K\geq1+2\log n$. Commonly imposed regularity conditions
would require that $H_{n}(\beta_{*})\rightarrow^{d}N(0,\Omega_{*})$
for some matrix $\Omega_{*}\in\RR^{L\times L}$. Hence, we obtain
\[
\sqrt{n}(\Acal_{K}(\hat{\beta},\hGamma)-\beta_{*})\rightarrow^{d}N(0,(\Gamma_{*}'\Gamma_{*})^{-1}\Gamma_{*}'\Omega_{*}\Gamma_{*}(\Gamma_{*}'\Gamma_{*})^{-1}).
\]

Moreover, Theorem \ref{thm: inference main} also provides a robustness
guarantee on the asymptotic approximation. Since we are taking a supreme
in (\ref{eq: main inference conclusion}), the approximation of $\Acal_{K}(\beta,\Gamma)-\beta_{*}$
by $-n^{-1/2}(\Gamma_{*}'\Gamma_{*})^{-1}\Gamma_{*}'H_{n}(\beta_{*})$
holds uniformly in $(K,\beta,\Gamma)$. This means this approximation
is robust to choices of $(K,\beta,\Gamma)$. For example, one can
run the $k$-step, update the initial estimates, use the output to
run the $k$-step iterations again, update the initial estimates,
and repeat these steps arbitrarily many times. Theorem \ref{thm: inference main}
says that by doing so, one should not expect to change the inference
results. We now summarize the entire procedure for estimation and
inference in Algorithm \ref{alg: final inference}.

\begin{algorithm}[h]
\caption{\label{alg: final inference}Estimation and inference for non-smooth
GMM}

\begin{raggedright}
Implement the following steps:
\par\end{raggedright}
\begin{enumerate}
\item Compute $\bar{\beta}$ for $\beta_{*}$ and $\hGamma$ for $\Gamma_{*}$.
\item Compute $\hbeta=\Acal_{K}(\bbeta,\hGamma)$ with $K=1+\left\lceil 2\log n\right\rceil $.
\item Use $\hbeta$ to obtain a new estimate $\tGamma$.
\item Compute $\tbeta=\Acal_{K}(\hbeta,\tGamma)$.
\item Compute the asymptotic variance $\hV=(\tGamma'\tGamma)^{-1}\tGamma'\hat{\Omega}\tGamma(\tGamma'\tGamma)^{-1}$,
where $\hat{\Omega}=n^{-1}\sum_{i=1}^{n}g(W_{i};\tbeta)g(W_{i};\tbeta)'$.
\item Conduct inference for $\beta_{*}$ based on $\sqrt{n}\hV^{-1/2}(\tbeta-\beta_{*})\rightarrow^{d}N(0,I_{p})$.
\end{enumerate}
\end{algorithm}

In many empirical applications, the sample size $n$ can be enormous.
Notice that in Algorithm \ref{alg: final inference}, all the steps
require only matrix multiplication, except Step 1 and potentially
Step 3. In obtaining the initial estimation in Step 1, the MILP formulation
presented in Section \ref{sec: IVQR} would be computationally very
costly when the sample size $n$ exceeds 1000. However, since we do
not require the consistency of $\bbeta$ in Step 1, one can simply
run the MILP on a randomly selected subsample of size $m$, where
$m\ll n$. In simulations, we find that for $n=5\times10^{6}$, using
$m=500$ yields decent performance. In this case, we only use $m/n=0.01\%$
of the data for initial estimation and Algorithm \ref{alg: final inference}
takes less than 15 seconds! This is a massive reduction in computing
time because even linear programs can be slow in such massive scale.
Hence, Algorithm \ref{alg: final inference} can be used for large-scale
quantile regressions.%

\section{\label{sec: Jacobian estimate}Tuning-free Jacobian estimation}

The $k$-step correction in Section \ref{sec: k step} requires an
estimate for the Jacobian of the population moment condition. We now
provide a tuning-free option. To fix ideas, we recall $G(\beta)=Eg(W_{i};\beta)$.
Now we are interested in estimating the Jacobian
\[
\Gamma_{0}:=\Gamma(b_{0}),
\]
where $\Gamma(\beta)=\frac{\partial G(\beta)}{\partial\beta'}\in\RR^{L\times p}$
and $b_{0}$ is an observed quantity, either random (e.g., GMM estimator
or any initial estimate) or a deterministic quantity.
\begin{example}[IVQR Jacobian]
\label{exa: IVQR}For IVQR, recall the moment function $g(W_{i};\beta)=Z_{i}\left(\mathbf{1}\{Y_{i}\le X_{i}'\beta\}-\tau\right)$.
Then 
\[
\Gamma(\beta)=Ef(X_{i}'\beta|X_{i},Z_{i})Z_{i}X_{i}',
\]
where $f(a|b,c)$ is the conditional density of $Y_{i}$ at $a$ given
$(X_{i},Z_{i})=(b,c)$.
\end{example}
\begin{example}[Density estimation]
\label{exa: density estimation}For density estimation (indexed by
quantile), we consider 
\[
g(W_{i};\beta)=\oneb\{Y_{i}\le\beta\}.
\]
Then $\Gamma(\beta)$ is just the density of $Y_{i}$ at $\beta$.
\end{example}
In these examples, estimation of $\Gamma(\beta)$ is typically done
by a kernel method that requires a choice of bandwidth. Unfortunately
in many applications, the choice of bandwidth can be quite difficult
to determine and such a choice can often significantly affect the
final estimation and inference. Here, we propose a tuning-free estimation
scheme. Since the need of Jacobian estimation also arises in other
situations (e.g., estimating asymptotic variance), we think this proposal
is of independent interest. Moreover, for both examples, our proposal
can be easily impolemented by essentially closed-end formulas.

\subsection{Methodology and theory}

At a high level, the mechanism that delivers the tuning-free property
is closely related to the bootstrapped standard errors for quantile
regressions. There are two popular methods for computing the standard
errors for quantile regressions. One is to use the explicit formula
and replace unknown density in this formula with a kernel estimate,
which requires a bandwidth choice. The other is to bootstrap the quantile
estimate. Notice that the latter avoids explicitly choosing a tuning
parameter because the bootstrapped estimates naturally generate perturbations
in a local neighborhood that allow us to learn the slope; see Section
\ref{subsec: remark bs quantile} for more discussions. Here, we develop
a computationally simple scheme implementing a similar mechanism for
estimating the Jacobian for general GMM models. Our proposed estimator
for $\Gamma(\beta)$ is based on the following two observations:
\begin{enumerate}
\item $\Gamma(\beta)$ is a matrix of dimensional $L\times p$ and can be
estimated entry by entry. Hence, we only need to solve the one-dimensional
problem with $p=L=1$. In this section, we only consider this case.
\item The slope of $G(\beta)$ at $\beta$ can be explored by small deviations
around $\beta$. Instead of explicitly (e.g., in terms of bandwidth)
choosing the magnitude of these deviations, let us generate these
small deviations naturally via bootstrap-like resampling. 
\end{enumerate}
Let $\{\xi_{i}\}_{i=1}^{n}$ be random variables simulated independent
of the data with $E(\xi_{i})=1$, say i.i.d $N(1,1)$ or binary variables
in $\{0,2\}$ with equal probability. Recall $G_{n}(\beta)=n^{-1}\sum_{i=1}^{n}g(W_{i};\beta)$
and $H_{n}(\beta)=\sqrt{n}(G_{n}(\beta)-G(\beta))$. We define $G_{n}^{*}(\beta)=n^{-1}\sum_{i=1}^{n}\xi_{i}g(W_{i};\beta)$
and $H_{n}^{*}(\beta)=n^{-1/2}\sum_{i=1}^{n}(\xi_{i}-1)g(W_{i};\beta)$.
We have some flexibility in generating the multipliers; for example,
one can also simulate $(\xi_{1},...,\xi_{n})$ from a multinomial
distribution with parameter $n$ and probabilities $(n^{-1},...,n^{-1})$,
leading to the empirical bootstrap. 

To estimate $\Gamma(b_{0})$, we solve the one-dimensional problem
\begin{equation}
\text{find}\ b_{*}\in[b_{0}-\bC,b_{0}+\bC]\ s.t.\ G_{n}^{*}(b_{*})=G_{n}(b_{0}),\label{eq: num BS}
\end{equation}
where $\bC>0$ is a large enough constant. If there are multiple solutions,
we choose $b_{*}$ to be the solution closest to $b_{0}$. The following
result provides the basic intuition.
\begin{lem}
\label{lem: BS approx}Assume that $G(\cdot)$ is twice-continuously
differentiable. 
\begin{equation}
-n^{-1/2}H_{n}^{*}(b_{*})=\Gamma_{0}(b_{*}-b_{0})+\varepsilon,\label{eq: BS OLS}
\end{equation}
where $\varepsilon=a_{n}^{*}(b_{*}-b_{0})^{2}+n^{-1/2}\left(H_{n}(b_{*})-H_{n}(b_{0})\right)$
and $a_{n}^{*}$ is a random variable satisfying $P(|a_{n}^{*}|\leq\sup_{\beta}|d^{2}G(\beta)/d\beta^{2}|)=1$.
\end{lem}
Consider the case in which $G(\cdot)$ has a bounded second-order
derivative, $H_{n}(\cdot)$ is Donsker and $b_{*}-b_{0}=o_{P}(1)$.
Lemma \ref{lem: BS approx} essentially says
\[
-n^{-1/2}H_{n}^{*}(b_{*})\approx\Gamma_{0}\times(b_{*}-b_{0}).
\]

Since both $-n^{-1/2}H_{n}^{*}(b_{*})$ and $b_{*}-b_{0}$ are directly
observed, we can simulate enough of them and regress $-n^{-1/2}H_{n}^{*}(b_{*})$
on $b_{*}-b_{0}$ via OLS without intercept. Formally, we use simulation
to approximate
\begin{equation}
\hGamma_{0}=\frac{E\left(-n^{-1/2}H_{n}^{*}(b_{*})(b_{*}-b_{0})\mid\Fcal\right)}{E((b_{*}-b_{0})^{2}\mid\Fcal)},\label{eq: BS estimator}
\end{equation}
where $\Fcal$ is the $\sigma$-algebra generated by the data. We
now give the formal result.
\begin{thm}
\label{thm: BS Jacobian}Assume that the following hold:
\begin{enumerate}
\item There exist constants $\rho_{1},\rho_{2},\rho_{3},\rho_{4}>0$ such
that $\sup_{\beta}|d^{2}G(\beta)/d\beta^{2}|\leq\rho_{1}$, $P(\rho_{2}<|\Gamma_{0}|<\rho_{3})\rightarrow1$
and $P\left(E\left((H_{n}^{*}(b_{0}))^{2}\mid\Fcal\right)\geq\rho_{4}\right)\rightarrow1$. 
\item $\sup_{|x-y|\leq n^{-1/2}t}|H_{n}(x)-H_{n}(y)|=o_{P}(1)$ and $E\left(\sup_{|x-y|\leq n^{-1/2}t}|H_{n}^{*}(x)-H_{n}^{*}(y)|^{4}\mid\Fcal\right)=o_{P}(1)$
for any constant $t>0$.
\item $E\left(\sup_{\beta}|H_{n}^{*}(\beta)|^{4}\mid\Fcal\right)=O_{P}(1)$
and $\sup_{\beta}|H_{n}(\beta)|=O_{P}(1)$. 
\end{enumerate}
Then $|\hGamma_{0}-\Gamma_{0}|=o_{P}(1)$.
\end{thm}
In Theorem \ref{thm: BS Jacobian}, we allow $b_{0}$ to be random,
which means $\Gamma_{0}=\Gamma(b_{0})$ can be random; when $b_{0}$
is non-random, $P(\rho_{2}<|\Gamma_{0}|<\rho_{3})$ is either one
or zero. The assumptions on $H_{n}(\cdot)$ are satisfied when $H_{n}(\cdot)$
is Donsker. The conditions on $H_{n}^{*}(\cdot)$ can be easily verified
when $\{\xi_{i}\}_{i=1}^{n}$ are sub-Gaussian multipliers. In this
case, $H_{n}^{*}(\cdot)$ conditional on $\Fcal$ is a sub-Gaussian
process and one can invoke the usual maximal inequalities and tail
bounds in \citet{van1996weak}. Under these weak regularity conditions,
Theorem \ref{thm: BS Jacobian} establishes the consistency of $\hGamma_{0}$.
We now present simple computational methods for implementing this
estimation strategy.

\subsection{Computational aspects}

\subsubsection{Computation for Example \ref{exa: IVQR}}

In Example \ref{exa: IVQR}, the problem in (\ref{eq: num BS}) becomes
\[
\text{find}\ b_{*}\qquad s.t.\qquad\sum_{i=1}^{n}\mathbf{1}\{Y_{i}\leq X_{i}b_{*}\}Z_{i}\xi_{i}=d,
\]
where $d$ is a known number $d=\sum_{i=1}^{n}Z_{i}\left(\mathbf{1}\{Y_{i}\leq X_{i}b_{0}\}+(\xi_{i}-1)\tau\right)$.
Since $X_{i}$ and $Z_{i}$ are scalars, we can easily obtain a closed-end
solution. We summarize the idea below.
\begin{lem}
\label{lem: IVQR BS computation}Let $\{(y_{i},x_{i},w_{i})\}_{i=1}^{n}$
and $a$ be given real numbers. Assume that $x_{i}\neq0$ and $w_{i}\neq0$
$\forall1\leq i\leq n$. Let $b_{*}\in\RR$ be such that $y_{i}\neq x_{i}b_{*}$
$\forall1\leq i\leq n$. Then $b_{*}$ satisfies $\sum_{i=1}^{n}\mathbf{1}\{y_{i}\leq x_{i}b_{*}\}w_{i}=a$
if and only if $\sum_{i=1}^{n}\mathbf{1}\{\ty_{i}\leq b_{*}\}\tw_{i}=c$,
where $\ty_{i}=y_{i}/x_{i}$, $\tw_{i}=w_{i}\oneb\{i\in A_{+}\}-w_{i}\oneb\{i\in A_{-}\}$,
$c=a+\sum_{i\in A_{-}}\tw_{i}$, $A_{+}=\{i:\ x_{i}>0\}$ and $A_{-}=\{i:\ x_{i}<0\}$.
\end{lem}
By Lemma \ref{lem: IVQR BS computation}, we just need to use the
cumulative sum of $\tw_{i}$. We summarize the details in Algorithm
\ref{alg: compute BS IVQR}. To solve (\ref{eq: num BS}), we can
simply apply Algorithm \ref{alg: compute BS IVQR} with $a=\sum_{i=1}^{n}Z_{i}\left(\mathbf{1}\{Y_{i}\leq X_{i}\beta_{0}\}+(\xi_{i}-1)\tau\right)$
and obtain a solution that satisfies (\ref{eq: num BS}) with error
$O_{P}(n^{-1}\max_{1\leq i\leq n}|Z_{i}\xi_{i}|)$. 

\begin{algorithm}[h]
\caption{\label{alg: compute BS IVQR}Tuning-free Jacobian estimation for IVQR}

\begin{raggedright}
To find $b_{*}$ such that $\sum_{i=1}^{n}\mathbf{1}\{y_{i}\leq x_{i}b_{*}\}w_{i}=a$,
implement the following steps:
\par\end{raggedright}
\begin{enumerate}
\item Compute quantities $\ty_{i}$, $\tw_{i}$ and $c$ as in Lemma \ref{lem: IVQR BS computation}.
\item Sort the data by $\ty_{i}$, i.e., $\ty_{i}\leq\ty_{i+1}$.
\item Compute the cumulative sum $S_{j}=\sum_{i=1}^{j}\tw_{i}$.
\item Find $j_{*}=\arg\min_{1\leq j\leq n}|S_{j}-c|$
\item Choose $b_{*}$ from $\{\ty_{j_{*}}-\eta,\ty_{j_{*}}+\eta\}$, where
$\eta$ is a small number to break the tie favorably, e.g., $\eta=\min_{1\leq i\leq n-1}(\ty_{i+1}-\ty_{i})/2$.
\end{enumerate}
\end{algorithm}

One can also view Algorithm \ref{alg: compute BS IVQR} as a smart
grid since we are essentially saying that the solution has to be one
of $\{\ty_{i}\}_{i=1}^{n}$. Therefore, the number of grid points
and the location of the grid points are completely determined by the
data. Since the algorithm essentially only consists of sorting variables,
it can be implemented efficiently. In our experiment, for a sample
size $n=5102$, bootstrapping 2000 samples takes only 0.5 second.

\subsubsection{\label{subsec: remark bs quantile}Computation for Example \ref{exa: density estimation}
and connection to bootstrap quantiles}

We can use the method discussed above (with $X_{i}=Z_{i}=1$) for
the computation of Example \ref{exa: density estimation}. As mentioned
before, $\{\xi_{i}\}_{i=1}^{n}$ can be generated from a multinomial
distribution and in this case the computation procedure of $b_{*}$
described in Algorithm \ref{alg: compute BS IVQR} becomes computing
the empirical quantile in bootstrap samples. Therefore, our proposed
estimate $\hGamma_{0}$ has a natural connection with bootstrap standard
errors for quantile regression.

We would like to point out that bootstrapping quantile regressions
in this case can be viewed as an alternative application of Lemma
\ref{lem: BS approx}. In Example \ref{exa: density estimation},
we have$\Gamma_{0}=f_{Y}(b_{0})$, where $f_{Y}(\cdot)$ is the probability
density function of the scalar variable $Y_{i}$ and $b_{0}$ is the
$\tau$-quantile of the sample. Another way of writing (\ref{eq: BS OLS})
is $H_{n}^{*}(b_{*})\approx-\Gamma_{0}\sqrt{n}(b_{*}-b_{0})$. The
proposed $\hGamma_{0}$ regresses $H_{n}^{*}(b_{*})$ onto $\sqrt{n}(b_{*}-b_{0})$
without intercept. Alternatively, since $\Gamma_{0}$ is known to
be positive, we can simply estimate $\Gamma_{0}$ using $\sqrt{E[(H_{n}^{*}(b_{*}))^{2}\mid\Fcal]/E[n(b_{*}-b_{0})^{2}\mid\Fcal]}.$
Since $H_{n}^{*}(\cdot)$ is approximating a Brownian bridge, we know
that $E[(H_{n}^{*}(b_{*}))^{2}\mid\Fcal]\approx\tau(1-\tau)$ and
simplify this formula as 
\[
\sqrt{\frac{\tau(1-\tau)}{E[n(b_{*}-b_{0})^{2}\mid\Fcal]}}.
\]

Notice that this is exactly the density estimate that we implicitly
use when we use the bootstrap standard error. The asymptotic variance
formula for the sample quantile is $\tau(1-\tau)/\hat{f}_{Y}^{2}(b_{0})$
for some density estimate $\hf_{Y}(b_{0})$. If we equate this with
the bootstrap variance $E[n(b_{*}-b_{0})^{2}\mid\Fcal]$, we obtain
that the implicitly used $\hat{f}_{Y}(b_{0})$ is given by the above
formula. Therefore, bootstrapping sample quantiles can serve as a
tuning-free density estimator. Our estimator $\hGamma_{0}$, which
is an OLS estimator, extends this strategy to Jacobian estimation
for general GMM models. 

\section{\label{sec: IVQR}Initial estimator via mixed integer linear programming}

The methodology we have presented so far does not assume a particular
choice of the initial estimate. Therefore, one can choose from methods
in the existing literature. In this section, we provide an MILP approach,
which can be used for IVQR and related problems. We argue that this
is an attractive alternative since it capitalizes on recent advances
in mixed integer programming, which has been intensively studied over
the past decades and is starting to be applied in large-scale problems.
Here, we do not need to wait for MILP to find a global solution; theoretical
results on the statistical property of early termination are provided. 

\subsection{IVQR}

In this section, we consider the IV quantile model in (\ref{eq: linear IV quantile model}).
Our proposal is a method of moment approach:
\begin{equation}
\hat{\beta}=\arg\min_{\beta\in\mathcal{B}}\ \|E_{n}Z_{i}(\mathbf{1}\{Y_{i}-X_{i}'\beta\leq0\}-\tau)\|_{\infty},\label{eq: MILO low-dim}
\end{equation}
where $\mathcal{B}\subseteq\mathbb{R}^{p}$ is a convex set. In practice,
we can choose $\mathcal{B}=\mathbb{R}^{p}$ or a bounded rectangular
subset of $\mathbb{R}^{p}$. The above estimator is based on the fact
that $EZ_{i}(\mathbf{1}\{y_{i}-X_{i}'\beta\leq0\}-\tau)=0$ for $\beta=\beta_{*}$.
Of course we can replace $Z_{i}$ with transformations of $Z_{i}$.
The idea of the estimator is to find a value $\beta$ to minimize
the ``magnitude'' of the empirical version $E_{n}Z_{i}(\mathbf{1}\{y_{i}-X_{i}'\beta\leq0\}-\tau)$. 

The estimator (\ref{eq: MILO low-dim}) differs from GMM in that we
use the $\ell_{\infty}$-norm, instead of the $\ell_{2}$-norm. The
choice of $\ell_{\infty}$-norm over $\ell_{2}$-norm is due to computational
reasons. As we shall see, the formulation with $\ell_{\infty}$-norm
in (\ref{eq: MILO low-dim}) can be cast as an MILP. If we use $\ell_{2}$-norm
instead, then the optimization problem would become a mixed integer
quadratic program (MIQP), which is the formulation in \citet{chen2017exact}.\footnote{In their Appendix C3, an MILP formulation is provided, but it requires
much more binary variables. Their formulation needs $n+n(n\text{\textminus}1)/2$
binary variables, while our formulation requires $n$ binary variables.} However, as pointed out in \citet{hemmecke2010nonlinear,burer2012milp,mazumder2017discrete},
it is quite well known in the integer programming community that current
algorithms for MILP problems are a much more mature technology than
MIQP. For this reason, we use the formulations in (\ref{eq: MILO low-dim}).

\subsubsection{\label{subsec: formulate MILP for IVQR}Formulation as a mixed integer
linear program}

We now show that the estimator (\ref{eq: MILO low-dim}) can be cast
as an MILP. The key is to introduce $n$ binary variables and use
constraints to force them to represent $\mathbf{1}\{Y_{i}-X_{i}'\beta\leq0\}$.

Let $\xi_{i}\in\{0,1\}$. Suppose that $M>0$ is an arbitrary number
such that $\max_{1\leq i\leq n}|Y_{i}-X_{i}'\hat{\beta}|\leq M$.
Notice that this is not a statistical tuning parameter since we can
choose any large enough $M>0$. The key insight is to realize that
imposing the constraint $-M\xi_{i}\leq Y_{i}-X_{i}'\beta\leq M(1-\xi_{i})$
will force $\xi_{i}$ to behave like $\mathbf{1}\{Y_{i}-X_{i}'\beta\leq0\}$.
To see this, consider the following two cases (ignoring the case of
$Y_{i}-X_{i}'\beta=0$): (1) $Y_{i}-X_{i}'\beta<0$ and (2) $Y_{i}-X_{i}'\beta>0$.
In Case (1), $\xi_{i}=1$ is the only possibility to make $-M\xi_{i}\leq Y_{i}-X_{i}'\beta\leq M(1-\xi_{i})$
hold. Similarly, in Case (2), $\xi_{i}=0$ is the only choice of $\xi_{i}$
in $\{0,1\}$ to satisfy the constraint. Hence, we need to consider
variables $\xi_{i}\in\{0,1\}$ and $\beta\in\mathbb{R}^{p}$ such
that $-M\xi_{i}\leq Y_{i}-X_{i}'\beta\leq M(1-\xi_{i})$.

In order to minimize $\|E_{n}Z_{i}(\xi_{i}-\tau)\|_{\infty}$, we
introduce an auxiliary variable $t\geq0$ with the constraint $-t\leq E_{n}Z_{i,j}(\xi_{i}-\tau)\leq t$
for $j\in\{1,...,L\}$, where $Z_{i,j}$ is the $j$th component of
$Z_{i}$. By minimizing $t$, we equivalently achieve minimizing $\|E_{n}Z_{i}(\xi_{i}-\tau)\|_{\infty}$.
To summarize, the final MILP formulation reads 
\begin{eqnarray}
(\hat{\beta},\hat{\xi},\hat{t}) & = & \underset{(\beta,\xi,t)}{\arg\min}\ t\label{eq: MILP construction}\\
 & s.t. & -M\xi_{i}\leq Y_{i}-X_{i}'\beta\leq M(1-\xi_{i})\nonumber \\
 &  & -\mathbf{1}_{L}t\leq E_{n}Z_{i}(\xi_{i}-\tau)\leq\mathbf{1}_{L}t\nonumber \\
 &  & \xi_{i}\in\{0,1\},\ \beta\in\mathcal{B},\ t\geq0.\nonumber 
\end{eqnarray}

In the case of $Y_{i}-X_{i}'\beta=0$, we have an indeterminancy since
both $\xi_{i}=0$ and $\xi_{i}=1$ would satisfy $-M\xi_{i}\leq Y_{i}-X_{i}'\beta\leq M(1-\xi_{i})$.
However, for most of the design matrices, $\{i:\ Y_{i}-X_{i}'\beta=0\}$
is empty. If we encounter a lot of zeros for $Y_{i}-X_{i}'\beta$
in the solution, we can simply incorporate a small wedge to solve
the indeterminacy: $-M\xi_{i}+D\leq Y_{i}-X_{i}'\beta\leq M(1-\xi_{i})$,
where $D>0$ is a very small number, such as machine precision tolerance.
In our experience, this is not necessary and does not make a difference
in the solution. 

\subsubsection{\label{subsec: bound MILP}Bounding the estimation error under early
termination}

We now derive the rate of convergence of $\hat{\beta}$. We also discuss
how the rate is affected if we terminate MILP before a global solution
is reached. A practical guide for early termination is provided and
its theoretical validity is also established. 

We start with the following simple high-level condition for identification.
Let us introduce the following notations. Recall the notations $G(\beta)=EZ_{i}(\mathbf{1}\{Y_{i}-X_{i}'\beta\leq0\}-\tau)$,
$G_{n}(\beta)=n^{-1}\sum_{i=1}^{n}Z_{i}(\mathbf{1}\{Y_{i}-X_{i}'\beta\leq0\}-\tau)$
and $H_{n}(\beta)=\sqrt{n}(G_{n}(\beta)-G(\beta))$.
\begin{assumption}
\label{assu: ID}Suppose that $\beta_{*}\in\Bcal$. For any $\eta>0$,
there exists a constant $C_{\eta}>0$ such that $\min_{\|\beta-\beta_{*}\|_{2}\geq\eta}\|G(\beta)\|_{2}\geq C_{\eta}$.
Moreover, there exist constants $c_{1},c_{2}>0$ such that 
\[
\inf_{\|\beta-\beta_{*}\|_{2}\leq c_{1}}\frac{\|G(\beta)\|_{2}}{\|\beta-\beta_{*}\|_{2}}\geq c_{2}.
\]
\end{assumption}
Assumption \ref{assu: ID} guarantees the identification of $\beta_{*}$
and can be verified using primitive conditions similar to Assumption
2 in \citet{chernozhukov2006instrumental}. In this paper, we do not
consider the case with weak identification.\footnote{Inference under potentially weak instruments is quite challenging
even for linear IV models. For joint inference on the entire vector
$\beta$ or all the coefficients of the endogenous variables, we can
rely on the method proposed in \citet{Chernozhukov2008}. However,
for subvector inference (inference only on part of endogenous variables),
it is quite challenging even in the linear IV models, for which weak
identification has been thoroughly understood only for  homoscedastic
errors; see e.g., \citet{guggenberger2012asymptotic}. } We also assume that the empirical process for $Z_{i}(\mathbf{1}\{Y_{i}-X_{i}'\beta\leq0\}-\tau)$
is globally Glivenko-Cantelli and locally Donsker. 
\begin{assumption}
\label{assu: Donsker} Suppose that $\sup_{\beta\in\mathcal{B}}\|n^{-1/2}H_{n}(\beta)\|_{2}=o_{P}(1)$.
Moreover, there exists a constant $c>0$ such that $\sup_{\|\beta-\beta_{*}\|_{2}\leq c}\|H_{n}(\beta)\|_{2}=O_{P}(1)$.
\end{assumption}
Assumption \ref{assu: Donsker} is not difficult to verify. For example,
straight-forward arguments using Lemmas 2.6.15 and 2.6.18 in \citet{van1996weak}
imply that under enough moments of $\|Z_{i}\|_{2}$, the entropy condition
in Theorem 2.14.1 therein holds, which means that $E\sup_{\|\beta-\beta_{*}\|_{2}\leq c}\|H_{n}(\beta)\|_{2}=O(1)$.
Since we typically terminate the MILP algorithm before a global solution
is found, we would like to consider the properties of estimations
from early termination. 
\begin{thm}
\label{thm: rate of convergence}Let Assumptions \ref{assu: ID} and
\ref{assu: Donsker} hold. Let $\hbeta\in\RR^{p}$ be an estimator.
If $\|\hG_{n}(\hbeta)\|_{\infty}=o_{P}(1)$, then $\|\hbeta-\beta_{*}\|_{2}\leq O_{P}(\|G_{n}(\hbeta)\|_{\infty}+n^{-1/2})$. 
\end{thm}
Theorem \ref{thm: rate of convergence} says that when $\|G_{n}(\hbeta)\|_{\infty}$
is small, the rate for $\|\hbeta-\beta_{*}\|_{2}$ is $\|G_{n}(\hbeta)\|_{\infty}+n^{-1/2}$.
Notice that we observe $\|G_{n}(\hbeta)\|_{\infty}$ in the MILP algorithm.
Hence, we can terminate it once it reaches certain threshold. A natural
threshold is $\|G_{n}(\beta_{*})\|_{\infty}$. Although we cannot
really compute $\|G_{n}(\beta_{*})\|_{\infty}$ in practice, we can
provide a finite-sample bound for it using the moderate deviation
result for self-normalized sums. Let $Z_{1,j}$ denote the $j$-th
component of $Z_{i}\in\RR^{L}$. 
\begin{lem}
\label{lem: bnd self normaliz}Suppose that there exist constants
$\xi_{1},\xi_{2}>0$ such that $\max_{1\leq j\leq L}E|Z_{1,j}|^{3}\left|\mathbf{1}\{\varepsilon_{i}\leq0\}-\tau\right|^{3}\le\xi_{1}$
and $\min_{1\leq j\leq L}EZ_{1,j}^{2}\left(\mathbf{1}\{\varepsilon_{i}\leq0\}-\tau\right)^{2}\geq\xi_{2}$.
Then there exists a constant $C>0$ depending only on $\xi_{1},\xi_{2}$
such that for any $n\geq C$ and any $\alpha\geq1/n$, 
\[
P\left(\|G_{n}(\beta_{*})\|_{\infty}>\Phi^{-1}(1-\alpha/n)n^{-1}\sqrt{\max_{1\leq j\leq L}\sum_{i=1}^{n}Z_{i,j}^{2}}\right)\leq4L\alpha n^{-1}.
\]
\end{lem}
In practice, we can simply take $\alpha=1/n$ and thus Lemma \ref{lem: bnd self normaliz}
tells us that for $n$ not too small, we have 
\[
P\left(\|G_{n}(\beta_{*})\|_{\infty}>Q_{*}\right)\leq4Ln^{-2},
\]
where $Q_{*}=\Phi^{-1}(1-n^{-2})n^{-1}\sqrt{\max_{1\leq j\leq L}\sum_{i=1}^{n}Z_{i,j}^{2}}$.
Notice that $Q_{*}$ can be explicitly computed from the data. Moreover,
we know that $Q_{*}=O_{P}(\sqrt{n^{-1}\log n})$. Therefore, if we
stop the MILP algorithm once $\|G_{n}(\hbeta)\|_{\infty}\leq Q_{*}$,
Lemma \ref{lem: bnd self normaliz} and Theorem \ref{thm: rate of convergence}
imply that $\|\hbeta-\beta_{*}\|_{2}=O_{P}(\sqrt{n^{-1}\log n})$.
As we have seen in Section \ref{sec: k step}, this is more than enough
for the $k$-step correction to yield an estimator that is asymptotically
equivalent to GMM. 

Now we provide simulation results to illustrate this point. We find
that the MILP algorithm reaches $Q_{*}$ within seconds. Let $p=20$.
We generate $Y_{i}=X_{i}'\theta+(X_{i}'\gamma)U_{i}$, where $X_{i}$
and $U_{i}$ are generated from the uniform distribution on $(0,1)$.
Entries of $\theta$ and $\gamma$ are randomly generated from the
uniform distribution on $(0,1)$. We set $\tau=0.7$. The starting
point of the MILP algorithm is generated from $N(0,I_{p})$. In Table
\ref{tab: early termination}, we report the frequency of $\|G_{n}(\hbeta)\|_{\infty}\leq Q_{*}$
based on 1000 simulations. 

\begin{table}[h]
\caption{\label{tab: early termination}Frequency of $\|G_{n}(\protect\hbeta)\|_{\infty}\protect\leq Q_{*}$
with early termination of MILP}

\bigskip{}

\begin{centering}
\bigskip{}
\par\end{centering}
\begin{centering}
\begin{tabular}{cccc}
 &  &  & \tabularnewline
$P\left(\|G_{n}(\hbeta)\|_{\infty}\leq Q_{*}\right)$ & $Z=X$ & $Z=\log X$ & $Z=[X,\log X]$\tabularnewline
\hline 
$n=200$, $t=5$ & 1.0000 & 0.9990 & 1.0000\tabularnewline
$n=500$, $t=5$ & 0.9990 & 0.9970 & 0.9970\tabularnewline
$n=500$, $t=10$ & 1.0000 & 1.0000 & 0.9980\tabularnewline
 &  &  & \tabularnewline
\end{tabular}
\par\end{centering}
{\small{}The following table shows $P\left(\|G_{n}(\hbeta)\|_{\infty}\leq Q_{*}\right)$,
where $\hbeta$ is obtained by terminating MILP after $t$ seconds
and $Q_{*}=\Phi^{-1}(1-n^{-2})n^{-1}\sqrt{\max_{1\leq j\leq L}\sum_{i=1}^{n}Z_{i,j}^{2}}$.}{\small\par}
\end{table}

As we can see from Table \ref{tab: early termination}, we only need
to run the algorithm for 10 seconds to ensure that $\|G_{n}(\hbeta)\|_{\infty}\leq Q_{*}$,
which implies $\|\hbeta-\beta_{*}\|_{2}=O_{P}(\sqrt{n^{-1}\log n})$.

\subsection{High-dimensional IV quantile regression}

When $p\gg n$ and $\beta$ is a sparse vector, the model (\ref{eq: linear IV quantile model})
becomes a high-dimensional IV quantile model. Although our $k$-step
correction framework does not cover the case of $p$ growing with
$n$, we still present the formulation of estimating high-dimensional
IV quantile regression to demonstrate the generality of MILP. In high
dimensions, successful estimation relies on proper regularization
on $\beta$. Similar to the regularization in Dantzig selector for
linear models (\citet{candes2007dantzig}), we propose 
\begin{eqnarray*}
\hat{\beta} & = & \arg\min_{\beta}\ \|\beta\|_{1}\\
 & s.t. & \left\Vert E_{n}Z_{i}\left(\mathbf{1}\{Y_{i}-X_{i}'\beta\leq0\}-\tau\right)\right\Vert _{\infty}\leq\lambda,
\end{eqnarray*}
where $\lambda\asymp\sqrt{n^{-1}\log p}$ is tuning parameter. 

Similar to the formulation in Section \ref{sec: IVQR}, we can cast
the above problem as an MILP. To account for the $\ell_{1}$-norm
in the objective function, we decompose each entry of $\beta$ into
the positive and negative part: we write $\beta_{j}=\beta_{j}^{+}-\beta_{j}^{-}$
with $\beta_{j}^{+},\beta_{j}^{-}\geq0$. Then the above problem can
be rewritten as
\begin{eqnarray}
\hat{\beta} & = & \underset{\beta^{+},\beta^{-}\in\mathbb{R}^{p},\ \xi=(\xi_{1},...,\xi_{n})\in\mathbb{R}^{n}}{\arg\min}\quad\sum_{j=1}^{p}\beta_{j}^{+}+\sum_{j=1}^{p}\beta_{j}^{-1}\label{eq: MILO}\\
 & s.t. & -\lambda\mathbf{1}_{L}\leq E_{n}Z_{i}\left(\xi_{i}-\tau\right)\leq\lambda\mathbf{1}_{L}\nonumber \\
 &  & -M\xi_{i}\leq y_{i}-X_{i}'(\beta^{+}-\beta^{-})\leq M(1-\xi_{i})\nonumber \\
 &  & \xi_{i}\in\{0,1\}\nonumber \\
 &  & \beta_{j}^{+},\beta_{j}^{-}\geq0.\nonumber 
\end{eqnarray}

\subsection{\label{subsec: censored regression}Censored regressions}

The censored regression proposed by \citet{powell1986censored} reads
\begin{equation}
\hat{\theta}=\underset{\theta\in\mathbb{R}^{p}}{\arg\min}\ E_{n}\rho_{\tau}\left(Y_{i}-\max\{X_{i}'\theta,0\}\right),\label{eq: censored reg}
\end{equation}
where $\rho_{\tau}(x)=x(\tau-\mathbf{1}\{x\leq0\})$ is the ``check''
function for a given $\tau\in(0,1)$ and $\{(Y_{i},X_{i})\}_{i=1}^{n}$
is the observed data. Notice that this is a nonconvex and non-smooth
optimization problem. Computationally it might not be very attractive,
especially when the dimensionality is large. The literature has seen
alternative estimators that explicitly model the probability of being
censored; see e.g., \citet{buchinsky1998alternative,chernozhukov2002three}.
Recently, there is work in high-dimensional statistics (e.g., \citet{muller2016censored})
studying the statistical properties of 
\begin{equation}
\hat{\theta}=\underset{\theta\in\mathbb{R}^{p}}{\arg\min}\ E_{n}\rho_{\tau}\left(Y_{i}-\max\{X_{i}'\theta,0\}\right)+\lambda\|\theta\|_{1},\label{eq: HD censored reg}
\end{equation}
where $\lambda\asymp\sqrt{n^{-1}\log p}$ is a tuning parameter. However,
discussions regarding the computational burden for the above estimator
are not common. Here, we case the problem (\ref{eq: HD censored reg})
as a MILP. Since problem (\ref{eq: censored reg}) is a special case
of problem (\ref{eq: HD censored reg}) with $\lambda=0$, our framework
can be used for the computation of both (\ref{eq: censored reg})
and (\ref{eq: HD censored reg}).

We introduce variables $\zeta_{i}^{+},\zeta_{i}^{-}\geq0$ to denote
the positive and negative parts of $Y_{i}-\max\{X_{i}'\theta,0\}$:
$Y_{i}-\max\{X_{i}'\theta,0\}=\zeta_{i}^{+}-\zeta_{i}^{-}$. Similarly,
we introduce $r_{i}^{+},r_{i}^{-}\geq0$ such that $X_{i}'\theta=r_{i}^{+}-r_{i}^{-}$;
also, let $\theta_{j}^{+},\theta_{j}^{-}\geq0$ satisfy $\theta_{j}=\theta_{j}^{+}-\theta_{j}^{-}$.
As in Section \ref{sec: IVQR}, we use $\xi_{i}\in\{0,1\}$ to represent
$\mathbf{1}\{X_{i}'\theta<0\}$ by imposing $-\xi_{i}M\leq X_{i}'\theta\leq(1-\xi_{i})M$,
where $M>0$ is any number satisfying $\|X\hat{\theta}\|_{\infty}\leq M$. 

Notice that $\max\{X_{i}'\theta,0\}=r_{i}^{+}$ if we can force one
of $r_{i}^{+}$ and $r_{i}^{-}$ to be exactly zero. The key idea
to achieve this is to impose $0\leq r_{i}^{+}\leq M(1-\xi_{i})$ and
$0\leq r_{i}^{-}\leq M\xi_{i}$. If $X_{i}'\theta>0$, then $\xi_{i}=0$,
which forces $r_{i}^{-}=0$; if $X_{i}'\theta<0$, then $\xi_{i}=1$,
which forces $r_{i}^{+}=0$. Now we write down the MILP formulation
for (\ref{eq: HD censored reg}): 
\begin{eqnarray*}
 & \underset{r_{i}^{+},r_{i}^{-},\zeta_{i}^{+},\zeta_{i}^{-},\theta_{j}^{+},\theta_{j}^{-},\xi_{i}}{\arg\min} & \frac{\tau}{n}\sum_{i=1}^{n}\zeta_{i}^{+}+\frac{1-\tau}{n}\sum_{i=1}^{n}\zeta_{i}^{-}+\lambda\sum_{j=1}^{p}\theta_{j}^{+}+\lambda\sum_{j=1}^{p}\theta_{j}^{-}\\
 & s.t. & Y_{i}-r_{i}^{+}=\zeta_{i}^{+}-\zeta_{i}^{-}\\
 &  & -\xi_{i}M\leq X_{i}'(\theta^{+}-\theta^{-})\leq(1-\xi_{i})M\\
 &  & X_{i}'(\theta^{+}-\theta^{-})=r_{i}^{+}-r_{i}^{-}\\
 &  & 0\leq r_{i}^{+}\leq M(1-\xi_{i})\\
 &  & 0\leq r_{i}^{-}\leq M\xi_{i}\\
 &  & r_{i}^{+},r_{i}^{-},\zeta_{i}^{+},\zeta_{i}^{-},\theta_{j}^{+},\theta_{j}^{-}\geq0\\
 &  & \xi_{i}\in\{0,1\}.
\end{eqnarray*}

\subsection{Censored IV quantile regressions}

Consider the following moment condition: 
\[
P\left(Y_{i}\leq\max\{X_{i}'\beta,C_{i}\}\mid Z_{i}\right)=\tau,
\]
where we observe i.i.d $\{(Y_{i},X_{i},Z_{i},C_{i})\}_{i=1}^{n}$.
\citet{chernozhukov2015quantile} proposed an estimator strategy that
uses a control variable. Here, we consider a direct approach based
on the above moment condition: 
\begin{equation}
\hat{\beta}=\underset{\beta}{\arg\min}\|E_{n}Z_{i}\left(\mathbf{1}\{Y_{i}\leq\max\{X_{i}'\beta,C_{i}\}\}-\tau\right)\|_{\infty}\label{eq: censored IV QR}
\end{equation}

Now we rewrite (\ref{eq: censored IV QR}) as an MILP. Similar to
Section \ref{subsec: censored regression}, we shall introduce binary
variables for the max function. Then we use additional binary variables
for the indicator function. 

We start by introducing $r_{i}^{+},r_{i}^{-}\geq0$ and $\xi_{i}\in\{0,1\}$
such that $X_{i}'\beta-C_{i}=r_{i}^{+}-r_{i}^{-}$, $-\xi_{i}M\leq X_{i}'\beta-C_{i}\leq(1-\xi_{i})M$,
$0\leq r_{i}^{+}\leq M(1-\xi_{i})$ and $0\leq r_{i}^{-}\leq M\xi_{i}$,
where $M>0$ is a large enough number. As explained in Section \ref{subsec: censored regression},
these constraints will force $\xi_{i}$ to behave like $\mathbf{1}\{X_{i}'\beta<C_{i}\}$
and ensure that one of $r_{i}^{+}$ and $r_{i}^{-}$ is exactly zero,
thus $r_{i}^{+}=\max\{X_{i}'\beta-C_{i},0\}$. Hence, $Y_{i}-\max\{X_{i}'\beta,C_{i}\}\leq0$
becomes $Y_{i}-C_{i}-r_{i}^{+}\leq0$. 

Now we introduce $q_{i}\in\{0,1\}$ such that $-Mq_{i}\leq Y_{i}-C_{i}-r_{i}^{+}\leq M(1-q_{i})$.
Again, this constraint would would make $q_{i}$ behave like $\mathbf{1}\{Y_{i}-C_{i}-r_{i}^{+}\leq0\}$.
Therefore, we only need to introduce an extra variable $t\geq0$ to
serve as $\|E_{n}Z_{i}(\xi_{i}-\tau)\|_{\infty}$. The final formulation
reads 
\begin{eqnarray*}
 & \underset{r_{i}^{+},r_{i}^{-},\xi_{i},q=(q_{1},...,q_{n})',t}{\arg\min} & t\\
 & s.t. & X_{i}'\beta-C_{i}=r_{i}^{+}-r_{i}^{-}\\
 &  & -\xi_{i}M\leq X_{i}'\beta-C_{i}\leq(1-\xi_{i})M\\
 &  & 0\leq r_{i}^{+}\leq M(1-\xi_{i})\\
 &  & 0\leq r_{i}^{-}\leq M\xi_{i}\\
 &  & -Mq_{i}\leq Y_{i}-C_{i}-r_{i}^{+}\leq M(1-q_{i})\\
 &  & -\mathbf{1}_{L}nt\leq Z'(q-\mathbf{1}_{n}\tau)\leq\mathbf{1}_{L}nt\\
 &  & r_{i}^{+},r_{i}^{-},t\geq0\\
 &  & \xi_{i},q_{i}\in\{0,1\}
\end{eqnarray*}

\section{\label{sec:Monte-Carlo-simulations}Monte Carlo simulations}

\subsection{\label{subsec: MC low-dim}Simulations for IVQR}

We now conduct simulations for Algorithm \ref{alg: final inference}
and set $m=500$, . As discussed above, when we run the MILP on the
subsample of size $m$, we can expect the rate of convergence to be
$\sqrt{m^{-1}\log m}$. We report the coverage probabilities of 95\%
confidence intervals for $\beta_{j}(\tau)$ for $1\leq j\leq p$. 

The results provide quite favorable evidence for the proposed estimator.
The empirical coverage probability is close to the nominal level of
confidence intervals. This is quite impressive for large $n$. When
$n=5\times10^{6}$ and $m=500$, we only use $0.01\%$ of the data
for MILP. This still yields good performance in terms of coverage
probability of confidence intervals.

We generate 
\[
Y_{i}=\alpha_{0}+\alpha D_{i}+\sum_{j=1}^{q}W_{i,j}\gamma_{j}+\sum_{j=1}^{q}D_{i}W_{i,j}\theta_{j}+\left(2\sqrt{3}q+\sum_{j=1}^{q}W_{i,j}\lambda_{j}+\sum_{j=1}^{q}D_{i}W_{i,j}\pi_{j}\right)V_{i},
\]
where $W_{i,j}\sim\text{uniform}(-\sqrt{3},\sqrt{3})$, $V_{i}\sim N(0,1)$
and $(D_{i},S_{i})$ are mutually independent. We generate $(S_{i},D_{i})\in\{(1,1),(1,0),(0,0)\}$
with probability 0.42, 0.25 and 0.33; these frequencies are estimated
from the JTPA data. We set $\alpha_{0}=\alpha=\gamma_{j}=\theta_{j}=\lambda_{j}=\pi_{j}=1$
and $q=10$. We set $X_{i}=(1,D_{i},W_{i,1},...,W_{i,q},D_{i}W_{i,1},...,D_{i}W_{i,q})\in\RR^{p}$
and $Z_{i}=(1,S_{i},W_{i,1},...,W_{i,q},S_{i}W_{i,1},...,S_{i}W_{i,q})'\in\RR^{p}$
with $p=2q+2=22$. One can easily verify that for any $\tau\in(0,1)$,
\[
P\left(Y_{i}\leq X_{i}'\beta(\tau)\mid Z_{i}\right)=\tau,
\]
with $\beta(\tau)=(1+2\sqrt{3}+c_{\tau},1+c_{\tau},1+c_{\tau},...,1+c_{\tau})'\in\RR^{p}$
and $c_{\tau}$ is the $\tau$-th quantile of $N(0,1)$. 

We use Gurobi 8.1 for mixed integer programming and implement it in
Matlab R2019a. For the starting values, we first randomly generate
$\beta_{{\rm start}}$ from $N(0,I_{p})$ and compute the starting
points for $\xi_{i}$ and $t$ using $\mathbf{1}\{Y_{i}-X_{i}'\beta_{{\rm start}}\leq0\}$
and $\|E_{n}Z_{i}(\mathbf{1}\{Y_{i}-X_{i}'\beta_{{\rm start}}\leq0\}-\tau)\|_{\infty}$,
respectively. We run the MILP program on a randomly selected subsample
of size $m=500$ and terminate the optimization algorithm after 5
seconds although a strict guarantee for global solutions would typically
take a few hours. The output of the MILP program is used as the initial
estimate for the $k$-step iteration. We report the performance of
the $k$-step corrected estimator in Table \ref{tab:RMSE}, which
is based on 800 random samples. In the experiments, we choose $n\in\{5\times10^{3},5\times10^{4},10^{6}\}$,
which means $m/n\in\{10\%,1\%,0.05\%\}$. Five quantiles $\tau\in\{0.15,0.25,0.5,0.75,0.85\}$
are considered. Since the usual $\chi^{2}$-test involves inverting
the $p\times p$ asymptotic variance matrix, the finite-sample performance
of the test might not be ideal as $p$ increases. One main reason
is that the asymptotic variance matrix is estimated and could be badly
conditioned even if $p$ is only 20. For this reason, we consider
the following non-pivotal test statistic that avoids inverting a large
matrix. 
\begin{rem}
\label{rem: box CS}Suppose that we have derived $\sqrt{n}(\hbeta-\beta)\rightarrow^{d}N(0,V)$
and computed $\hV$ as an estimator for $\hV$. Testing $H_{0}:\ \beta=\beta_{0}$
can be done using the Wald-test statistic $n(\hbeta-\beta_{0})'\hV^{-1}(\hbeta-\beta_{0})$
with the pivotal limiting distribution of $\chi^{2}(p)$. Another
option is to use $\sqrt{n}\|\hbeta-\beta\|_{\infty}$ as the test
statistic and compute $\Phi_{\alpha}(\hV^{1/2})$ as the critical
value, where $\Phi_{\alpha}(A)\in\RR$ denotes the number satisfying
$P(\|A\xi\|_{\infty}>\Phi_{\alpha}(A))=\alpha$ with $\xi\sim N(0,I_{p})$
for any matrix $A$. The computation of $\Phi_{\alpha}(\hV^{1/2})$
is very fast via simulation.\footnote{There might be another reason for why one would expect this non-pivotal
test to perform better for large $p$. We typically can derive that
$\hbeta-\beta\approx n^{-1}\sum_{i=1}^{n}\psi_{i}$ with $\psi_{i}\in\RR^{p}$
having mean zero. Currently, it is known that Gaussian approximation
can be more easily verified under the $\ell_{\infty}$-norm than the
$\ell_{2}$-norm. By results in \citet{chernozhukov2013gaussian},
Gaussian approximation holds for $\|n^{-1/2}\sum_{i=1}^{n}\psi_{i}\|_{\infty}$
even if $p\gg n$; in contrast, as far as we known, the best result
for Gaussian approximation of $\|n^{-1/2}\sum_{i=1}^{n}\psi_{i}\|_{2}$
requires $p\ll n^{1/4}$, see \citet{pouzo2015}.} In the rest of the paper, confidence sets based on inverting the
Wald test and the above non-pivotal test will be referred to as the
ellipsoid and rectangle confidence set, respectively. 
\end{rem}
In Table \ref{tab:RMSE}, we consider inference of $\beta(\tau)=(\beta_{1}(\tau),...,\beta_{p}(\tau))'$.
Notice that $\beta_{2}(\tau)$ corresponds to the coefficient for
$D_{i}$, $(\beta_{3}(\tau),...,\beta_{2+q}(\tau))$ for $(W_{i,1},...,W_{i,q})$
and $(\beta_{3+q}(\tau),...,\beta_{p}(\tau))$ for $(D_{i}W_{i,1},...,D_{i}W_{i,q})$.
We consider the coverage probability of confidence intervals (sets)
for various components (and subvectors) of $\beta(\tau)$. We see
that when the sample size is 5000, ellipsoid confidence sets, which
requires inverting a $22\times22$ estimated asymptotic variance,
have some undercoverage while rectangular confidence sets still have
accurate coverage probabilities. For larger sample size, all the confidence
intervals (sets) have quite accurate performance. We include the results
for $n=10^{6}$ to emphasize the point that for extremely large samples,
our method is still quite fast as it takes less than 15 seconds to
compute one sample. In this case, the main factor that limits the
speed is the memory. 
\begin{center}
\begin{table}[h]
\caption{\label{tab:RMSE}Inference of IVQR with $p=22$}

\begin{centering}
\begin{tabular}{lcccccccccccc}
 &  &  &  &  &  &  &  &  &  &  &  & \tabularnewline
 & \multicolumn{11}{c}{{\small{}$n=5000$ (for $k$-step) and $m=500$ (for MILP)}} & \tabularnewline
 & \multicolumn{5}{c}{{\small{}95\% confidence intervals (sets)}} &  & \multicolumn{5}{c}{{\small{}90\% confidence intervals (sets)}} & \tabularnewline
{\small{}$\text{Parameter }\backslash\qquad\tau$} & {\small{}$0.15$} & {\small{}$0.25$} & {\small{}0.5} & {\small{}0.75} & {\small{}0.85} &  & {\small{}$0.15$} & {\small{}$0.25$} & {\small{}0.5} & {\small{}0.75} & {\small{}0.85} & \tabularnewline
\cline{1-12} \cline{2-12} \cline{3-12} \cline{4-12} \cline{5-12} \cline{6-12} \cline{7-12} \cline{8-12} \cline{9-12} \cline{10-12} \cline{11-12} \cline{12-12} 
{\small{}$\beta_{2}(\tau)$} & {\small{}0.941} & {\small{}0.950} & {\small{}0.954} & {\small{}0.940} & {\small{}0.954} &  & {\small{}0.884} & {\small{}0.888} & {\small{}0.914} & {\small{}0.903} & {\small{}0.891} & \tabularnewline
{\small{}$\beta_{3}(\tau)$} & {\small{}0.958} & {\small{}0.953} & {\small{}0.949} & {\small{}0.949} & {\small{}0.950} &  & {\small{}0.913} & {\small{}0.893} & {\small{}0.896} & {\small{}0.909} & {\small{}0.904} & \tabularnewline
{\small{}$\beta_{3+q}(\tau)$} & {\small{}0.943} & {\small{}0.946} & {\small{}0.948} & {\small{}0.945} & {\small{}0.945} &  & {\small{}0.898} & {\small{}0.900} & {\small{}0.893} & {\small{}0.896} & {\small{}0.888} & \tabularnewline
{\small{}$\beta(\tau)$ (Ellipsoid)} & {\small{}0.860} & {\small{}0.909} & {\small{}0.933} & {\small{}0.894} & {\small{}0.866} &  & {\small{}0.783} & {\small{}0.851} & {\small{}0.883} & {\small{}0.818} & {\small{}0.773} & \tabularnewline
{\small{}$\beta_{3:(2+q)}(\tau)$ (Ellipsoid)} & {\small{}0.954} & {\small{}0.949} & {\small{}0.975} & {\small{}0.949} & {\small{}0.955} &  & {\small{}0.906} & {\small{}0.919} & {\small{}0.938} & {\small{}0.888} & {\small{}0.901} & \tabularnewline
{\small{}$\beta_{(3+q):p}(\tau)$ (Ellipsoid)} & {\small{}0.920} & {\small{}0.938} & {\small{}0.943} & {\small{}0.940} & {\small{}0.943} &  & {\small{}0.874} & {\small{}0.881} & {\small{}0.881} & {\small{}0.876} & {\small{}0.874} & \tabularnewline
{\small{}$\beta(\tau)$ (Rectangle)} & {\small{}0.951} & {\small{}0.965} & {\small{}0.955} & {\small{}0.950} & {\small{}0.969} &  & {\small{}0.909} & {\small{}0.919} & {\small{}0.900} & {\small{}0.903} & {\small{}0.904} & \tabularnewline
{\small{}$\beta_{3:(2+q)}(\tau)$ (Rectangle)} & {\small{}0.960} & {\small{}0.955} & {\small{}0.966} & {\small{}0.953} & {\small{}0.959} &  & {\small{}0.909} & {\small{}0.910} & {\small{}0.921} & {\small{}0.906} & {\small{}0.921} & \tabularnewline
{\small{}$\beta_{(3+q):p}(\tau)$ (Rectangle)} & {\small{}0.954} & {\small{}0.968} & {\small{}0.953} & {\small{}0.950} & {\small{}0.968} &  & {\small{}0.903} & {\small{}0.918} & {\small{}0.898} & {\small{}0.911} & {\small{}0.896} & \tabularnewline
 &  &  &  &  &  &  &  &  &  &  & \tabularnewline
 & \multicolumn{11}{c}{{\small{}$n=50,000$ (for $k$-step) and $m=500$ (for MILP)}} & \tabularnewline
 & \multicolumn{5}{c}{{\small{}95\% confidence intervals (sets)}} &  & \multicolumn{5}{c}{{\small{}90\% confidence intervals (sets)}} & \tabularnewline
{\small{}$\text{Parameter }\backslash\qquad\tau$} & {\small{}$0.15$} & {\small{}$0.25$} & {\small{}0.5} & {\small{}0.75} & {\small{}0.85} &  & {\small{}$0.15$} & {\small{}$0.25$} & {\small{}0.5} & {\small{}0.75} & {\small{}0.85} & \tabularnewline
\cline{1-12} \cline{2-12} \cline{3-12} \cline{4-12} \cline{5-12} \cline{6-12} \cline{7-12} \cline{8-12} \cline{9-12} \cline{10-12} \cline{11-12} \cline{12-12} 
{\small{}$\beta_{2}(\tau)$} & {\small{}0.948} & {\small{}0.950} & {\small{}0.949} & {\small{}0.948} & {\small{}0.958} &  & {\small{}0.889} & {\small{}0.895} & {\small{}0.895} & {\small{}0.889} & {\small{}0.885} & \tabularnewline
{\small{}$\beta_{3}(\tau)$} & {\small{}0.954} & {\small{}0.948} & {\small{}0.939} & {\small{}0.949} & {\small{}0.933} &  & {\small{}0.915} & {\small{}0.903} & {\small{}0.890} & {\small{}0.903} & {\small{}0.885} & \tabularnewline
{\small{}$\beta_{3+q}(\tau)$} & {\small{}0.941} & {\small{}0.960} & {\small{}0.935} & {\small{}0.935} & {\small{}0.958} &  & {\small{}0.884} & {\small{}0.913} & {\small{}0.871} & {\small{}0.888} & {\small{}0.909} & \tabularnewline
{\small{}$\beta(\tau)$ (Ellipsoid)} & {\small{}0.926} & {\small{}0.938} & {\small{}0.944} & {\small{}0.931} & {\small{}0.938} &  & {\small{}0.866} & {\small{}0.898} & {\small{}0.903} & {\small{}0.893} & {\small{}0.870} & \tabularnewline
{\small{}$\beta_{3:(2+q)}(\tau)$ (Ellipsoid)} & {\small{}0.954} & {\small{}0.946} & {\small{}0.946} & {\small{}0.955} & {\small{}0.943} &  & {\small{}0.901} & {\small{}0.884} & {\small{}0.894} & {\small{}0.908} & {\small{}0.886} & \tabularnewline
{\small{}$\beta_{(3+q):p}(\tau)$ (Ellipsoid)} & {\small{}0.940} & {\small{}0.949} & {\small{}0.943} & {\small{}0.948} & {\small{}0.949} &  & {\small{}0.880} & {\small{}0.896} & {\small{}0.893} & {\small{}0.890} & {\small{}0.894} & \tabularnewline
{\small{}$\beta(\tau)$ (Rectangle)} & {\small{}0.941} & {\small{}0.944} & {\small{}0.950} & {\small{}0.949} & {\small{}0.958} &  & {\small{}0.904} & {\small{}0.896} & {\small{}0.909} & {\small{}0.894} & {\small{}0.895} & \tabularnewline
{\small{}$\beta_{3:(2+q)}(\tau)$ (Rectangle)} & {\small{}0.945} & {\small{}0.950} & {\small{}0.956} & {\small{}0.959} & {\small{}0.940} &  & {\small{}0.909} & {\small{}0.885} & {\small{}0.913} & {\small{}0.911} & {\small{}0.896} & \tabularnewline
{\small{}$\beta_{(3+q):p}(\tau)$ (Rectangle)} & {\small{}0.941} & {\small{}0.945} & {\small{}0.953} & {\small{}0.950} & {\small{}0.958} &  & {\small{}0.898} & {\small{}0.898} & {\small{}0.911} & {\small{}0.890} & {\small{}0.896} & \tabularnewline
 &  &  &  &  &  &  &  &  &  &  &  & \tabularnewline
 & \multicolumn{11}{c}{{\small{}$n=1\ \text{million}$ (for $k$-step) and $m=500$ (for
MILP)}} & \tabularnewline
 & \multicolumn{5}{c}{{\small{}95\% confidence intervals (sets)}} &  & \multicolumn{5}{c}{{\small{}90\% confidence intervals (sets)}} & \tabularnewline
{\small{}$\text{Parameter }\backslash\qquad\tau$} & {\small{}$0.15$} & {\small{}$0.25$} & {\small{}0.5} & {\small{}0.75} & {\small{}0.85} &  & {\small{}$0.15$} & {\small{}$0.25$} & {\small{}0.5} & {\small{}0.75} & {\small{}0.85} & \tabularnewline
\cline{1-12} \cline{2-12} \cline{3-12} \cline{4-12} \cline{5-12} \cline{6-12} \cline{7-12} \cline{8-12} \cline{9-12} \cline{10-12} \cline{11-12} \cline{12-12} 
{\small{}$\beta_{2}(\tau)$} & {\small{}0.956} & {\small{}0.945} & {\small{}0.960} & {\small{}0.954} & {\small{}0.949} &  & {\small{}0.916} & {\small{}0.895} & {\small{}0.884} & {\small{}0.896} & {\small{}0.903} & \tabularnewline
{\small{}$\beta_{3}(\tau)$} & {\small{}0.938} & {\small{}0.945} & {\small{}0.961} & {\small{}0.936} & {\small{}0.946} &  & {\small{}0.903} & {\small{}0.914} & {\small{}0.919} & {\small{}0.883} & {\small{}0.889} & \tabularnewline
{\small{}$\beta_{3+q}(\tau)$} & {\small{}0.956} & {\small{}0.956} & {\small{}0.948} & {\small{}0.939} & {\small{}0.951} &  & {\small{}0.916} & {\small{}0.908} & {\small{}0.906} & {\small{}0.874} & {\small{}0.895} & \tabularnewline
{\small{}$\beta(\tau)$ (Ellipsoid)} & {\small{}0.940} & {\small{}0.941} & {\small{}0.961} & {\small{}0.941} & {\small{}0.956} &  & {\small{}0.885} & {\small{}0.894} & {\small{}0.905} & {\small{}0.894} & {\small{}0.898} & \tabularnewline
{\small{}$\beta_{3:(2+q)}(\tau)$ (Ellipsoid)} & {\small{}0.940} & {\small{}0.948} & {\small{}0.955} & {\small{}0.945} & {\small{}0.941} &  & {\small{}0.880} & {\small{}0.889} & {\small{}0.903} & {\small{}0.890} & {\small{}0.885} & \tabularnewline
{\small{}$\beta_{(3+q):p}(\tau)$ (Ellipsoid)} & {\small{}0.943} & {\small{}0.944} & {\small{}0.955} & {\small{}0.948} & {\small{}0.946} &  & {\small{}0.901} & {\small{}0.888} & {\small{}0.914} & {\small{}0.904} & {\small{}0.909} & \tabularnewline
{\small{}$\beta(\tau)$ (Rectangle)} & {\small{}0.938} & {\small{}0.943} & {\small{}0.948} & {\small{}0.951} & {\small{}0.965} &  & {\small{}0.883} & {\small{}0.878} & {\small{}0.893} & {\small{}0.896} & {\small{}0.908} & \tabularnewline
{\small{}$\beta_{3:(2+q)}(\tau)$ (Rectangle)} & {\small{}0.943} & {\small{}0.938} & {\small{}0.949} & {\small{}0.950} & {\small{}0.944} &  & {\small{}0.888} & {\small{}0.893} & {\small{}0.900} & {\small{}0.898} & {\small{}0.890} & \tabularnewline
{\small{}$\beta_{(3+q):p}(\tau)$ (Rectangle)} & {\small{}0.931} & {\small{}0.944} & {\small{}0.948} & {\small{}0.951} & {\small{}0.963} &  & {\small{}0.886} & {\small{}0.884} & {\small{}0.894} & {\small{}0.893} & {\small{}0.901} & \tabularnewline
 &  &  &  &  &  &  &  &  &  &  &  & \tabularnewline
\end{tabular}
\par\end{centering}

\end{table}
\par\end{center}

\subsection{Simulations for tuning-free derivative estimation}

Consider $Y_{i}=X_{i}+Z_{i}\varepsilon_{i}$ and $X_{i}=Z_{i}V_{i}$,
where $\varepsilon_{i}\sim\text{Exp}(\lambda)$,\footnote{$\text{Exp}(\lambda)$ denotes the exponential distribution with parameter
$\lambda$. The mean of this distribution is $\lambda^{-1}$.} $V_{i}\sim{\rm uniform}(0,1)$ and $Z_{i}\sim{\rm uniform}(0,2)$
are mutually independent. Let $g(W_{i};\beta)=Z_{i}\oneb\{Y_{i}\leq X_{i}\beta\}$.
We can explicitly compute the population derivative: for $\beta>1$,
\[
\Gamma(\beta)=\frac{dEg(W_{i};\beta)}{d\beta}=\frac{1-[\lambda(\beta-1)+1]\exp\left(\lambda(1-\beta)\right)}{\lambda(\beta-1)^{2}}.
\]

In Table \ref{tab: BS Jocbian RMSE}, we evaluate the performance
of two estimators in terms of root-mean-squared error (RMSE). The
first estimator $\hGamma_{\text{tuning-free}}$ is the one proposed
in Section \ref{sec: Jacobian estimate} computed using $\sqrt{n}$
bootstrap samples. The second estimator $\hGamma_{\text{kernel}}$
is the kernel estimator with Gaussian kernel and bandwidth given by
Silverman's rule of thumb. We see that the tuning-free estimator is
a good alternative to well-tuned kernel estimators. Since theoretically
$\hGamma_{\text{kernel}}$ has a faster rate of convergence, we see
that the constants in the bandwidth choice can be quite important
for performance in practice. The tuning-free estimator is attractive
in that one does not have to make these tricky choices. 
\begin{center}
\begin{table}[h]
\caption{\label{tab: BS Jocbian RMSE}RMSE of estimating $\Gamma(\beta)$ (measured
in $10^{-2}$)}

\centering{}%
\begin{tabular}{lcccccccccccccc}
 &  &  &  &  &  &  &  &  &  &  &  &  &  & \tabularnewline
 & \multicolumn{5}{c}{$\lambda=10$, $\beta=1.5$} &  &  & \multicolumn{5}{c}{$\lambda=10$, $\beta=3$} &  & \tabularnewline
$n$ & 400 & 700 & 1000 & 1300 & 1600 &  &  & 400 & 700 & 1000 & 1300 & 1600 &  & \tabularnewline
$\hGamma_{\text{tuning-free}}$ & 9.286 & 7.595 & 6.802 & 6.264 & 5.888 &  &  & 1.139 & 0.887 & 0.748 & 0.674 & 0.622 &  & \tabularnewline
$\hGamma_{\text{kernel}}$ & 5.650 & 4.537 & 4.050 & 3.690 & 3.384 &  &  & 1.669 & 1.446 & 1.328 & 1.237 & 1.177 &  & \tabularnewline
 &  &  &  &  &  &  &  &  &  &  &  &  &  & \tabularnewline
 & \multicolumn{5}{c}{$\lambda=1/3$, $\beta=1.5$} &  &  & \multicolumn{5}{c}{$\lambda=1/3$, $\beta=3$} &  & \tabularnewline
$n$ & 400 & 700 & 1000 & 1300 & 1600 &  &  & 400 & 700 & 1000 & 1300 & 1600 &  & \tabularnewline
$\hGamma_{\text{tuning-free}}$ & 2.105 & 1.739 & 1.611 & 1.513 & 1.439 &  &  & 1.589 & 1.391 & 1.259 & 1.187 & 1.119 &  & \tabularnewline
$\hGamma_{\text{kernel}}$ & 7.015 & 6.646 & 6.392 & 6.211 & 6.048 &  &  & 2.018 & 1.727 & 1.540 & 1.427 & 1.341 &  & \tabularnewline
 &  &  &  &  &  &  &  &  &  &  &  &  &  & \tabularnewline
\end{tabular}
\end{table}
\par\end{center}

\section{\label{sec: Empirical-Illustration}Empirical Illustration: heterogeneous
returns to training}

In Section \ref{sec:Introduction}, we mentioned the problem of investigating
the effect of JTPA. We now provide more details. The participation
status will be denoted by $D_{i}\in\{0,1\}$, where $D_{i}=1$ means
that individual $i$ participates in the program. The random offers,
denoted by $S_{i}\in\{0,1\}$, will be used as instruments, where
$S_{i}=1$ means that individual $i$ has an offer to participate.
Following \citet{Chernozhukov2008}, we also consider other 13 exogenous
variables denoted by $\{W_{i,j}\}_{j=1}^{13}$.\footnote{The data is downloaded from Christian Hansen's \href{http://faculty.chicagobooth.edu/christian.hansen/research/sampdata.zip}{website}.}
The outcome variable $Y_{i}$ is earnings. We consider the following
model: 
\begin{equation}
P\left(Y_{i}\leq\alpha_{0}(\tau)+D_{i}\alpha(\tau)+\sum_{j=1}^{13}D_{i}W_{i,j}\theta_{j}(\tau)+\sum_{j=1}^{13}W_{i,j}\gamma_{j}(\tau)\mid S_{i},\{W_{i,j}\}_{j=1}^{13}\right)=\tau,\label{eq: jpta moment}
\end{equation}
where $\tau\in(0,1)$. Under our notation (\ref{eq: linear IV quantile model}),
we have 
\[
\begin{cases}
X_{i}=(1,D_{i},D_{i}W_{i,1},...,D_{i}W_{i,13},W_{i,1},...,W_{i,13})'\in\mathbb{R}^{p}\\
Z_{i}=(1,S_{i},S_{i}W_{i,1},...,S_{i}W_{i,13},W_{i,1},...,W_{i,13})'\in\mathbb{R}^{L},
\end{cases}
\]
where $p=L=28$. We rescale $Z_{i}$ such that $n^{-1}\sum_{i=1}^{n}Z_{i,2:28}Z_{i,2:28}'=I_{27}$
for $j\in\{1,...,L\}$, where $Z_{i,2:28}$ denotes the vector $Z_{i}$
with the first component removed. We are interested in $\alpha(\tau)$,
which denotes the overall effect of JTPA, as well as $\theta_{j}(\tau)$,
which measures the heterogeneity of the effect. Following \citet{Chernozhukov2008},
we consider $\tau\in\{0.15,0.25,0.5,0.75,0.85\}$. We use the proposed
methods to test the model parameters. In Table \ref{tab: JTPA 1},
we report the test statistics for three hypotheses at these 5 quantiles. 

We find strong evidence for heterogeneity in treatment effects. The
coefficient $\alpha(\tau)$ for the treatment status is not significant\footnote{Since we include interaction terms in (\ref{eq: jpta moment}), the
estimates for $\alpha(\tau)$ would be not represent the ``average''
effect if there is heterogeneity in the treatment effect; hence, these
results should not be directly comparable to results reported in \citet{abadie2002instrumental}
and \citet{Chernozhukov2008}.}; moreover, the controls by themselves are insignificant or barely
significant at 5\% or 10\% level. However, we reject the insignificance
of $\theta_{j}(\tau)$ coefficients, which means that the covariates
are very informative on the magnitude of treatmente effects. 
\begin{center}
\begin{table}[h]
\caption{\label{tab: JTPA 1}Test statistics for JTPA}

\begin{centering}
\begin{tabular}{lr@{\extracolsep{0pt}.}lr@{\extracolsep{0pt}.}lr@{\extracolsep{0pt}.}lr@{\extracolsep{0pt}.}lr@{\extracolsep{0pt}.}l}
 & \multicolumn{2}{c}{} & \multicolumn{2}{c}{} & \multicolumn{2}{c}{} & \multicolumn{2}{c}{} & \multicolumn{2}{c}{}\tabularnewline
Test statistic & \multicolumn{2}{c}{$\tau=0.15$} & \multicolumn{2}{c}{$\tau=0.25$} & \multicolumn{2}{c}{$\tau=0.50$} & \multicolumn{2}{c}{$\tau=0.75$} & \multicolumn{2}{c}{$\tau=0.85$}\tabularnewline
$H_{0}:\ \alpha(\tau)=0$ & -0&002 & -0&088 & -0&649 & -0&491 & -0&393\tabularnewline
$H_{0}:\ \gamma_{1}(\tau)=\cdots=\gamma_{13}(\tau)=0$ & 0&141 & 5&195 & 5&111 & 23&725 & 10&557\tabularnewline
$H_{0}:\ \theta_{1}(\tau)=\cdots=\theta_{13}(\tau)=0$ & 0&566 & 166&096 & 124&992 & 120&062 & 91&505\tabularnewline
 & \multicolumn{2}{c}{} & \multicolumn{2}{c}{} & \multicolumn{2}{c}{} & \multicolumn{2}{c}{} & \multicolumn{2}{c}{}\tabularnewline
\end{tabular}
\par\end{centering}
{\small{}In the above table, t-statistic is reported for testing $H_{0}:\ \alpha(\tau)=0$.
For the other two hypotheses, we report the Wald statistic: $n\hat{\theta}'\hat{V}^{-1}\hat{\theta}$,
where $\htheta$ is the estimated parameters under testing and $\hat{V}$
denotes the estimated asymptotic variance. The 95\% quantile and 90\%
quantile of $\chi^{2}(13)$ are 22.3620 and 19.8119, respectively.}{\small\par}
\end{table}
\par\end{center}

\begin{figure}[h]
\caption{\label{fig: JTPA 1}Heterogeneous treatment effect of JTPA}

\begin{centering}
\includegraphics[scale=0.55]{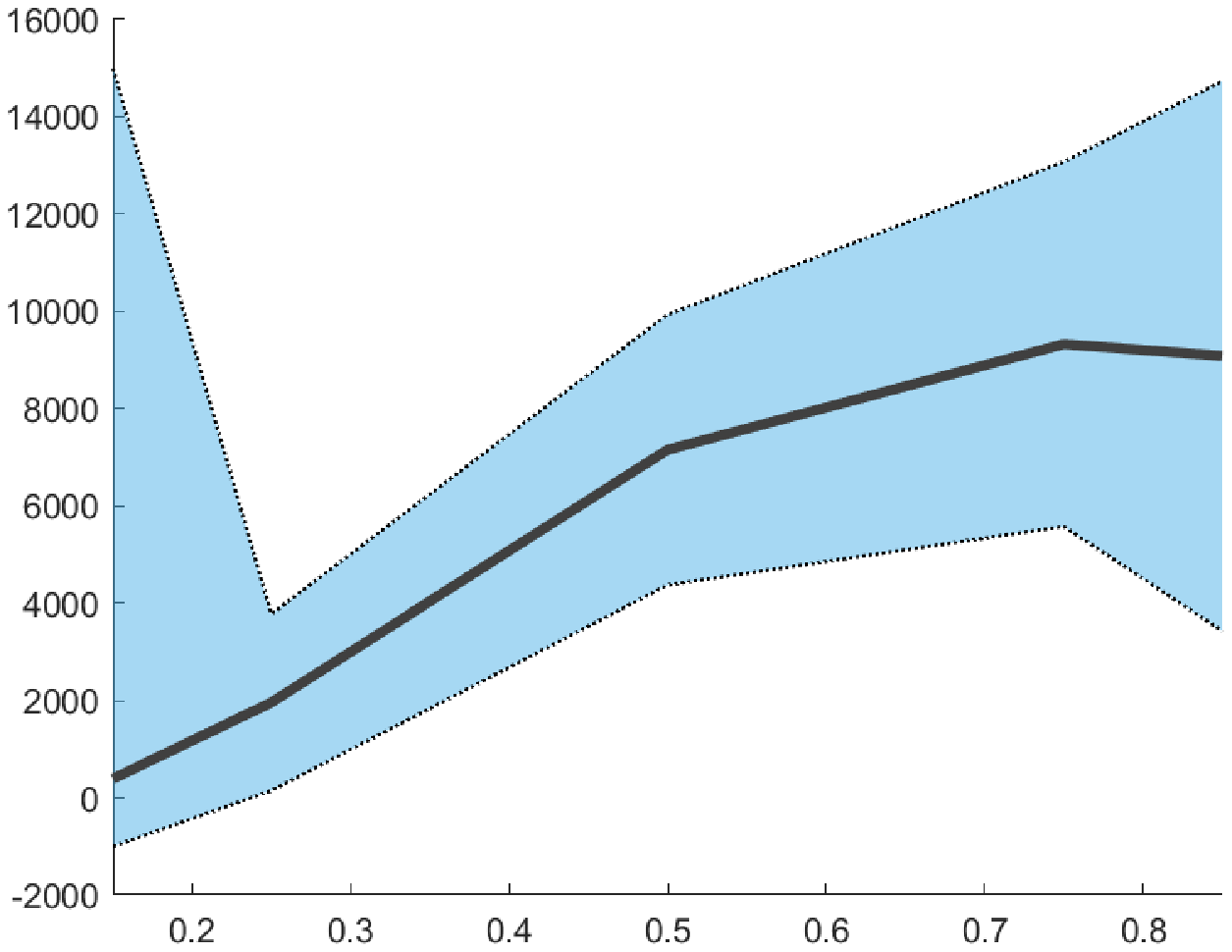}\includegraphics[scale=0.55]{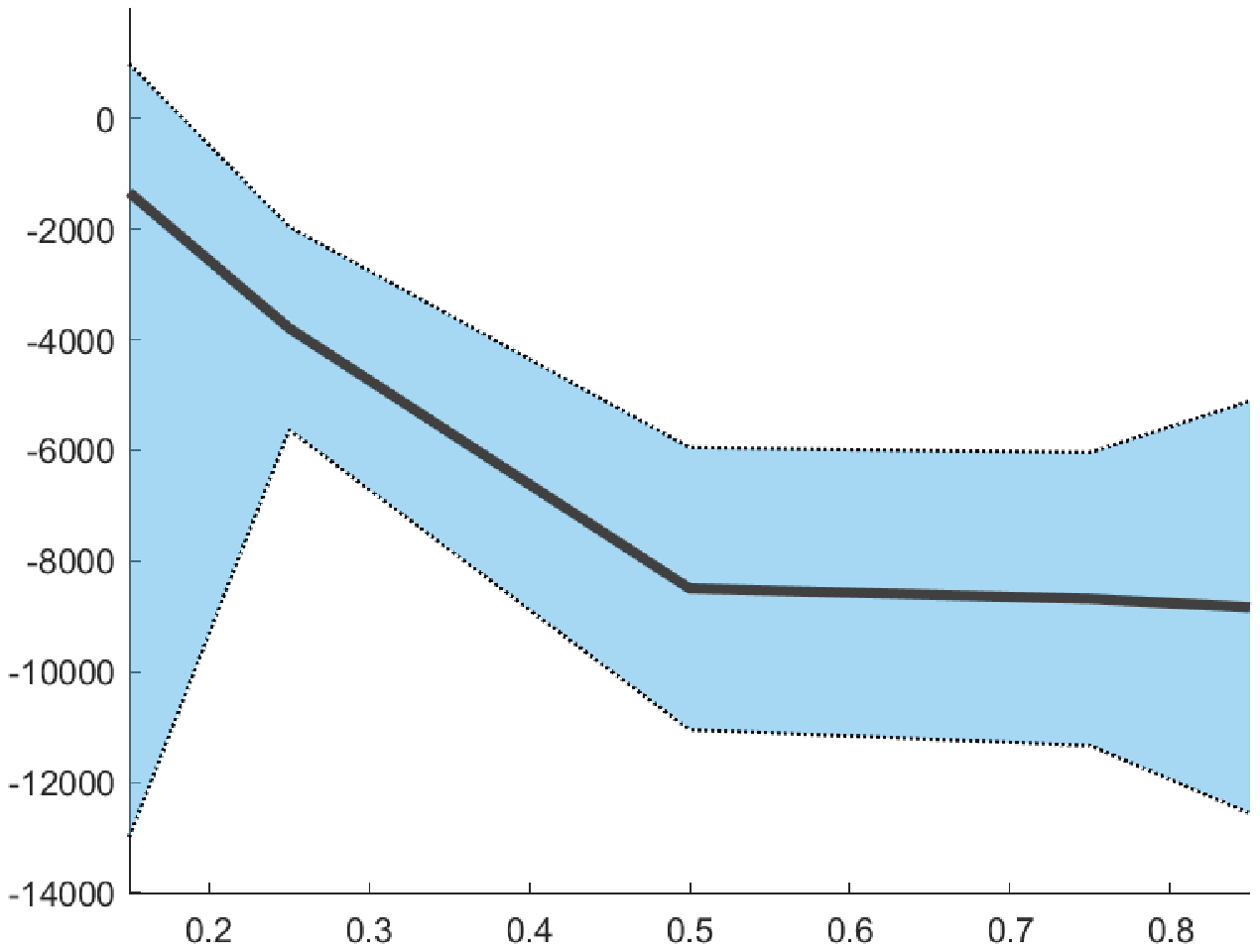}
\par\end{centering}
{\small{}The two figures plot the 95\% confidence bands for $\gamma_{4}(\tau)$
coefficient for $D_{i}\cdot W_{i,4}$ (left plot) and $\gamma_{5}(\tau)$
coefficient for $D_{i}\cdot W_{i,5}$ (right plot), where $W_{i,4}$
represents the marriage indicator (1 means being married) and $W_{i,5}$
represents the indicator of whether the person has worked for less
than 13 weeks in the past year.}{\small\par}
\end{figure}

We also present two prominent patterns in the heterogeneity. In Figure
\ref{fig: JTPA 1}, we see that the treatment effects of JTPA depend
on the marriage status and prior employment records. On the very left
tail, the the treatment effect does not depend on the marriage status,
whereas among higher-income individuals, married ones benefit more
from JTPA than the unmarried. Moreover, although prior employment
status is not related to the effect of JTPA for lower-income individuals,
we find that on the upper level of the income spectrum, those with
insufficient prior employment records tend to benefit less than those
with at least 13 weeks of employment in the past year. %

\section{Conclusion}

In this paper, we provide an alternative estimation and inference
method for IVQR and related problems. This alternative approach is
needed as we show that the GMM formulation of IVQR (or even a reasonable
approximation) is computationally NP-hard. Our proposal focuses on
obtaining good statistical properties instead of trying to improve
optimization algorithms for GMM. The proposed estimator can transform,
in a computationally efficient manner, an inconsistent initial estimator
into one that is asymptotically equivalent to GMM. The initial estimator
is obtained via MILP. We also propose a tuning-free method for estimating
the Jacobian of the moment condition in a non-smooth GMM model. We
illustrate our proposal in simulated and empirical data. %

\bibliographystyle{apalike}
\bibliography{SC_biblio}

\newpage{}

\appendix
\begin{center}
\textbf{\Large{}Appendix}{\Large\par}
\par\end{center}

\section{Proof of results in Section \ref{sec: k step}}
\begin{proof}[\textbf{Proof of Theorem \ref{thm: IVQR NP hard}}]
We consider the following slightly easier problem.
\begin{problem}
\label{prob: original IVQR V2}Given two numbers $\tau(0,1)$, $\eta\in(0,1/3)$
and data $\{(x_{i},z_{i},y_{i})\}_{i=1}^{n}$ with $x_{i}\in\RR^{p}$,
$z_{i}\in\RR^{L}$ and $y_{i}$, decide whether or not the following
holds:
\[
\min_{\beta\in\RR^{p}}\ \left\Vert \sum_{i=1}^{n}z_{i}\left(\oneb\{y_{i}-x_{i}'\beta\leq0\}-\tau\right)\right\Vert _{2}\leq\eta.
\]
\end{problem}
Problem \ref{prob: original IVQR V2} is easier than Problem \ref{prob: original IVQR}
because any algorithm that solves Problem \ref{prob: original IVQR}
can immediately tell us whether or not the optimal value of the objective
function is smaller than $\eta$. We show that Problem \ref{prob: original IVQR V2}
is NP-hard.

The rest of the proof follows the standard reduction principle. In
particular, we show that the partition problem, which is one of the
21 NP-complete problems presented in \citet{karp1972reducibility},
can be cast as Problem \ref{prob: original IVQR V2} (or reduced to
Problem \ref{prob: original IVQR V2})\footnote{Reduction to partition problem has been a useful tool for proving
NP-hardness of statistical procedures, see e.g., \citet{chen2014complexity}.}; see \citep{johnson1979computers,sipser2012introduction} for textbook
treatment on NP-completeness. Once we establish this reduction, the
proof is complete because such a reduction means that any efficient
algorithm solving Problem \ref{prob: original IVQR V2} is also an
efficient algorithm for solving the partition problem. Since the partition
problem is NP-hard, Problem \ref{prob: original IVQR V2} is also
NP-hard.
\begin{problem}[Partition]
\label{prob: subset sum}Given $m$ integers $\{a_{i}\}_{i=1}^{m}$,
decide whether or not the following holds: there exists a subset $B\subseteq\{1,...,m\}$
such that $\sum_{i\in B}a_{i}=\sum_{i\in\{1,...,m\}\backslash B}a_{i}$.
\end{problem}
Now we reduce Problem \ref{prob: subset sum} to Problem \ref{prob: original IVQR V2}.
Given $m$ integers $\{a_{i}\}_{i=1}^{m}$, we define $\tau=1/2$,
$\eta=1/4$, $p=L=n=m$, $y_{i}=1/2$. Using the notation $x_{i}=(x_{i,1},...,x_{i,p})'$
and $z_{i}=(z_{i,1},...,z_{i,p})'$, we set 
\[
x_{i,j}=\begin{cases}
\oneb\{i=j\} & i\leq m\\
0 & i>m
\end{cases}\qquad\text{and}\qquad z_{i,j}=\begin{cases}
a_{i} & i\leq m\\
0 & i>m
\end{cases}\qquad\forall1\le i\leq m.
\]
It now remains to prove the following claim.
\begin{claim*}
The answer to Problem \ref{prob: subset sum} is YES if and only if
the answer to Problem \ref{prob: original IVQR V2} is YES
\end{claim*}
\textbf{Step 1:} show the ``only if'' part.

Suppose that the answer to Problem \ref{prob: subset sum} is YES.
Then there exists $B\subset\{1,...,m\}$ such that $\sum_{i\in B}a_{i}=\sum_{i\in\{1,...,m\}\backslash B}a_{i}$.
Then simply choose $\beta=(\beta_{1},\beta_{2},...,\beta_{m})'$ with
$\beta_{j}=\oneb\{j\in B\}$. Then 
\[
y_{i}-x_{i}'\beta=\begin{cases}
-1/2 & \text{if }i\in B\\
1/2 & \text{if }i\in\{1,...,m\}\backslash B.
\end{cases}
\]

Then we notice that
\begin{align*}
 & \sum_{i=1}^{m}z_{i}\left(\oneb\{y_{i}-x_{i}'\beta\leq0\}-1/2\right)\\
 & =\sum_{i\in B}z_{i}\left(\oneb\{y_{i}-x_{i}'\beta\leq0\}-1/2\right)+\sum_{i\in\{1,...,m\}\backslash B}z_{i}\left(\oneb\{y_{i}-x_{i}'\beta\leq0\}-1/2\right)\\
 & =\sum_{i\in B}z_{i}\times(1/2)+\sum_{i\in\{1,...,m\}\backslash B}z_{i}\times(-1/2)\\
 & =\oneb_{L}\times\left(\sum_{i\in B}a_{i}-\sum_{i\in\{1,...,m\}\backslash B}a_{i}\right)\times(1/2)=0.
\end{align*}

Hence, $\|\sum_{i=1}^{m}z_{i}\left(\oneb\{y_{i}-x_{i}'\beta\leq0\}-1/2\right)\|_{2}=0\leq\eta$,
which means that the answer to Problem \ref{prob: original IVQR V2}
is YES.

\textbf{Step 2:} show the ``if'' part.

Suppose that the answer to Problem \ref{prob: original IVQR V2} is
YES. Then there exists $\beta\in\RR^{m}$ such that $\left\Vert \sum_{i=1}^{m}z_{i}\left(\oneb\{y_{i}-x_{i}'\beta\leq0\}-1/2\right)\right\Vert _{2}\leq\eta$.
Let $B=\{i:\ 1\leq i\leq m\ \text{and }x_{i}'\beta\geq1/2\}$.

Using the definition of $(y_{i},x_{i},z_{i})$, we know that 
\begin{align*}
 & 2\times\sum_{i=1}^{m}z_{i}\left(\oneb\{y_{i}-x_{i}'\beta\leq0\}-1/2\right)\\
 & =2\times\sum_{i\in B}z_{i}\left(\oneb\{x_{i}'\beta\geq1/2\}-1/2\right)+2\times\sum_{i\in\{1,...,m\}\backslash B}z_{i}\left(\oneb\{x_{i}'\beta\geq1/2\}-1/2\right)\\
 & =2\times\sum_{i\in B}z_{i}\times(1/2)+2\times\sum_{i\in\{1,...,m\}\backslash B}z_{i}\times(-1/2)=\sum_{i\in B}z_{i}-\sum_{i\in\{1,...,m\}\backslash B}z_{i}.
\end{align*}

Since $\left\Vert \sum_{i=1}^{m}z_{i}\left(\oneb\{y_{i}-x_{i}'\beta\leq0\}-1/2\right)\right\Vert _{2}\leq\eta$,
it follows that 
\[
\left\Vert \sum_{i\in B}z_{i}-\sum_{i\in\{1,...,m\}\backslash B}z_{i}\right\Vert _{2}=2\times\left\Vert \sum_{i=1}^{m}z_{i}\left(\oneb\{y_{i}-x_{i}'\beta\leq0\}-1/2\right)\right\Vert _{2}\leq2\eta,
\]
which means that 
\[
\left|\sum_{i\in B}z_{i,1}-\sum_{i\in\{1,...,m\}\backslash B}z_{i,1}\right|\leq2\eta.
\]

Since $\{z_{i,1}\}_{i=1}^{m}$ are integers $\{a_{i}\}_{i=1}^{m}$,
$\sum_{i\in B}z_{i,1}-\sum_{i\in\{1,...,m\}\backslash B}z_{i,1}$
is also an integer. By $2\eta<1$, it follows that
\[
\left|\sum_{i\in B}z_{i,1}-\sum_{i\in\{1,...,m\}\backslash B}z_{i,1}\right|=0,
\]
which means $\sum_{i\in B}z_{i,1}=\sum_{i\in\{1,...,m\}\backslash B}z_{i,1}$.
Since $z_{i,1}=a_{i}$ for $1\leq i\leq m$, the answer to Problem
\ref{prob: subset sum} is YES.

Notice that this same argument goes through if we replace $\|\cdot\|_{2}$-norm
with $\|\cdot\|_{r}$-norm.
\end{proof}
\begin{proof}[\textbf{Proof of Theorem \ref{thm: NP hard approx IVQR}}]
We first introduce the following minimum unsatisfied linear relations
(Min ULR). 
\begin{problem}[Min ULR]
\label{prob: minULR}Given $A\in\QQ^{n\times p}$ and $b\in\QQ^{n}$,
find $x\in\RR^{p}$ to minimize the number of violations in the relation
$Ax>b$, i.e., number of violations in the $n$ entrywise comparisons. 
\end{problem}
The following result due to \citet{amaldi1998approximability} says
that Min ULR cannot be approximated. 
\begin{thm}[Theorem 2 of \citet{amaldi1998approximability}]
\label{thm: approx NP hard min ULR}For any $\rho>1$, it is NP-hard
to find a $\rho$-approximation for Problem \ref{prob: minULR}. 
\end{thm}
We argue that Theorem \ref{thm: approx NP hard min ULR} still holds
even if we assume that $(A,b)\in(\QQ^{n\times p}\times\QQ^{n})\backslash\Kcal$
in Problem \ref{prob: minULR}, where $\Kcal=\{(A,b)\in\QQ^{n\times p}\times\QQ^{n}:\ Ax>b\ \text{for some }x\in\RR^{n}\}$.
To see this, suppose this is not true; in other words, for some $\rho>1$,
one can find a polynomial-time algorithm $\Acal$ to find a $\rho$-approximation
for all instances $(A,b)\in(\QQ^{n\times p}\times\QQ^{n})\backslash\Kcal$.
Then one can use the following polynomial-time algorithm to find $\rho$-approximations
for any instance $(A,b)\in\QQ^{n\times p}\times\QQ^{n}$. Start by
running the following linear program 
\begin{eqnarray*}
\underset{t\in\RR,\ x\in\RR^{n}}{\text{maximize}} & t\qquad\text{s.t.}\qquad A_{i}'x\geq t+b_{i} & \forall1\leq i\leq n,
\end{eqnarray*}
where $A_{i}'$ is the $i$th row of $A$ and $b_{i}$ the $i$th
component of $b$. If the solution is strictly positive, then stop;
otherwise, continue by running algorithm $\Acal$. This is a contradiction
to Theorem \ref{thm: approx NP hard min ULR} unless P=NP. Therefore,
Theorem \ref{thm: approx NP hard min ULR} still holds even if we
restrict Problem \ref{prob: minULR} to $(A,b)\in(\QQ^{n\times p}\times\QQ^{n})\backslash\Kcal$.
Hence, without loss of generality, we only consider instances in $(\QQ^{n\times p}\times\QQ^{n})\backslash\Kcal$. 

Now we cast Problem \ref{prob: minULR} as Problem \ref{prob: original IVQR}.
Let $(A,b)\in(\QQ^{n\times p}\times\QQ^{n})\backslash\Kcal$ be given.
We define $X=-A$ and $Y=-b$, $z_{i}=(1,...,1)'\in\RR^{L}$ and $\tau=1/(2n)$.
Let $x_{i}'$ denote the $i$th row of $X$ and $y_{i}$ the $i$th
component of $Y$. Notice that 
\begin{align*}
\left\Vert \sum_{i=1}^{n}z_{i}\left(\oneb\{y_{i}-x_{i}'\beta\leq0\}-\tau\right)\right\Vert _{2} & =\sqrt{L}\left|\sum_{i=1}^{n}\left(\oneb\{y_{i}-x_{i}'\beta\leq0\}-\tau\right)\right|\\
 & =\sqrt{L}\left|\sum_{i=1}^{n}\oneb\{y_{i}-x_{i}'\beta\leq0\}-\frac{1}{2}\right|\\
 & \overset{\text{(i)}}{\geq}\frac{\sqrt{L}}{2}\sum_{i=1}^{n}\oneb\{y_{i}-x_{i}'\beta\leq0\},
\end{align*}
where (i) follows by the fact that $(A,b)\in(\QQ^{n\times p}\times\QQ^{n})\backslash\Kcal$
(so $\sum_{i=1}^{n}\oneb\{y_{i}-x_{i}'\beta\leq0\}\geq1$). On the
other hand, we clearly have 
\begin{align*}
\left\Vert \sum_{i=1}^{n}Z_{i}\left(\oneb\{y_{i}-x_{i}'\beta\leq0\}-\tau\right)\right\Vert _{2} & =\sqrt{L}\left|\sum_{i=1}^{n}\oneb\{y_{i}-x_{i}'\beta\leq0\}-\frac{1}{2}\right|\\
 & \leq\sqrt{L}\sum_{i=1}^{n}\oneb\{y_{i}-x_{i}'\beta\leq0\}.
\end{align*}

Hence,
\[
\sqrt{L}/2\leq\frac{\left\Vert \sum_{i=1}^{n}Z_{i}\left(\oneb\{y_{i}-x_{i}'\beta\leq0\}-\tau\right)\right\Vert _{2}}{\sum_{i=1}^{n}\oneb\{y_{i}-x_{i}'\beta\leq0\}}\leq\sqrt{L}.
\]

Notice that $\sum_{i=1}^{n}\oneb\{y_{i}-x_{i}'\beta\leq0\}$ is equal
to the number of violations in $Ax>b$ with $x=\beta$. The above
display implies that any $\rho$-approximation for Problem \ref{prob: original IVQR}
can be used to construct a $2\rho$-approximation for Problem \ref{prob: minULR}.
By Theorem \ref{thm: approx NP hard min ULR}, the desired result
follows. Clearly, the above argument still holds when we replace $\|\cdot\|_{2}$
with $\|\cdot\|_{r}$. 
\end{proof}
\begin{proof}[\textbf{Proof of Lemma \ref{lem: main iter}}]
We observe that 
\begin{align}
 & \Acal(\beta,\hGamma)-\beta_{*}\nonumber \\
 & =(\beta-\beta_{*})-(\hGamma'\hGamma)^{-1}\hGamma'\left[G(\beta)+n^{-1/2}H_{n}(\beta)\right]\nonumber \\
 & =(\beta-\beta_{*})-(\hGamma'\hGamma)^{-1}\hGamma'\left[\Gamma_{*}(\beta-\beta_{*})+n^{-1/2}H_{n}(\beta)+\left(G(\beta)-\Gamma_{*}(\beta-\beta_{*})\right)\right]\nonumber \\
 & =(I-(\hGamma'\hGamma)^{-1}\hGamma'\Gamma_{*})(\beta-\beta_{*})-(\hGamma'\hGamma)^{-1}\hGamma'\left[n^{-1/2}H_{n}(\beta)+\left(G(\beta)-\Gamma_{*}(\beta-\beta_{*})\right)\right].\label{eq: key debias decomposition}
\end{align}

Notice that
\[
\|(I-(\hGamma'\hGamma)^{-1}\hGamma'\Gamma_{*})(\beta-\beta_{*})\|_{2}=\|(\hGamma'\hGamma)^{-1}\hGamma'(\hGamma-\Gamma_{*})(\beta-\beta_{*})\|_{2}\leq\|(\hGamma'\hGamma)^{-1}\hGamma\|\times\|\hGamma-\Gamma_{*}\|\times\|\beta-\beta_{*}\|_{2}.
\]

Since $\|G(\beta)-\Gamma_{*}(\beta-\beta_{*})\|_{2}\leq c\|\beta-\beta_{*}\|_{2}^{2}$
and $\|H_{n}(\beta)\|_{2}\leq\sup_{v\in\Bcal_{0}}\|H_{n}(v)\|_{2}$,
the desired result follows. 
\end{proof}

\begin{proof}[\textbf{Proof of Theorem \ref{thm: main iter rate}}]
First notice that the assumption of $c_{3}^{-1/2}c_{5}n^{-1/2}\leq(1-\rho_{*})c_{1}$
means that $\rho_{*}\leq1-n^{-1/2}c_{1}^{-1}c_{3}^{-1/2}c_{5}<1$. 

Notice that $\|(\hGamma'\hGamma)^{-1}\hGamma'\|_{2}=1/\sqrt{\lambda_{\min}(\hGamma'\hGamma)}\leq c_{3}^{-1/2}$.
Then by Lemma \ref{lem: main iter}, we have that 
\begin{align*}
\|\hbeta_{(1)}-\beta_{*}\|_{2} & \leq c_{3}^{-1/2}c_{2}\|\bbeta-\beta_{*}\|_{2}+c_{3}^{-1/2}\left(n^{-1/2}c_{5}+c_{4}\|\bbeta-\beta_{*}\|_{2}^{2}\right)\\
 & \leq c_{3}^{-1/2}c_{2}\|\bbeta-\beta_{*}\|_{2}+c_{3}^{-1/2}\left(n^{-1/2}c_{5}+c_{1}c_{4}\|\bbeta-\beta_{*}\|_{2}\right)\\
 & =\rho_{*}\|\bbeta-\beta_{*}\|_{2}+c_{3}^{-1/2}c_{5}n^{-1/2}.
\end{align*}

By the assumption of $c_{5}\leq\sqrt{n}c_{1}c_{3}^{1/2}(1-\rho_{*})$,
we have that $c_{3}^{-1/2}c_{5}n^{-1/2}\leq(1-\rho_{*})c_{1}$. Hence,
$\|\hbeta_{(1)}-\beta_{*}\|_{2}\leq c_{1}$. By induction, we have
that $\|\hbeta_{(k)}-\beta_{*}\|_{2}\leq c_{1}$ for any $k\geq1$. 

We now notice that the same computation as above yields that for any
$k\geq1$, 
\[
\|\hbeta_{(k)}-\beta_{*}\|_{2}\leq\rho_{*}\|\hbeta_{(k-1)}-\beta_{*}\|_{2}+c_{3}^{-1/2}c_{5}n^{-1/2}.
\]

Thus, the desired result follows by a simple induction argument.
\end{proof}

\begin{proof}[\textbf{Proof of Corollary \ref{cor: main rate}}]
We shall use the notations from the statement of Theorem \ref{thm: main iter rate}.
We also write $\hbeta_{(k)}=\Acal_{k}(\bbeta,\hGamma)$. Since Theorem
\ref{thm: main iter rate} is a finite-sample result that holds with
probability one, we can define $c_{1},...,c_{5}$ to be random or
non-random quantities. Set $c_{1}=\min\{\sqrt{\kappa_{3}}/(8\kappa_{1}),\ \kappa_{2}\}$,
$c_{2}=\sqrt{\kappa_{3}}/8$, $c_{3}=\kappa_{3}/4$, $c_{4}=\kappa_{1}$
and $c_{5}=\sup_{\|v-\beta_{*}\|_{2}\leq\kappa_{2}}\|H_{n}(v)\|_{2}$.
Let $\rho_{*}=c_{3}^{-1/2}(c_{2}+c_{1}c_{4})\leq1/2$. Define the
event 
\[
\Mcal=\left\{ \|\bbeta-\beta_{*}\|_{2}\leq c_{1}\right\} \bigcap\left\{ \|\hGamma-\Gamma_{*}\|\leq c_{2}\right\} \bigcap\left\{ \sup_{\|v-\beta_{*}\|_{2}\leq\kappa_{2}}\|H_{n}(v)\|_{2}\leq\sqrt{n}\kappa_{3}/(32\kappa_{1})\right\} .
\]

Now we verify the conditions of Theorem \ref{thm: main iter rate}
on the event $\Mcal$. 

Let $\sigma_{\min}(\cdot)$ and $\sigma_{\max}(\cdot)$ denote the
minimum and the maximum singular values, respectively. Then $\sigma_{\min}(\Gamma_{*})\geq\sqrt{\kappa_{3}}$.
Recall the elementary inequality of $\sigma_{\min}(A)+\sigma_{\max}(B)\geq\sigma_{\min}(A+B)$
for any matrices $A,B$. Hence, on the event $\Mcal$, 
\[
\sqrt{\lambda_{\min}(\hGamma'\hGamma)}=\sigma_{\min}(\hGamma)\geq\sigma_{\min}(\Gamma_{*})-\sigma_{\max}(\Gamma_{*}-\hGamma)=\sqrt{\kappa_{3}}-\|\hGamma-\Gamma\|\geq\sqrt{\kappa_{3}}-c_{2}=7\sqrt{\kappa_{3}}/8.
\]

Therefore, on the event $\Mcal$, $\lambda_{\min}(\hGamma'\hGamma)\geq\kappa_{3}(7/8)^{2}>c_{3}$.
Notice that on the event $\Mcal$, $c_{5}\leq c_{1}c_{3}^{1/2}\sqrt{n}(1-\rho_{*})$
by definition since $\rho_{*}\leq1/2$ and $c_{1}\leq\sqrt{\kappa_{3}}/(8\kappa_{1})$.
Therefore, all the conditions of Theorem \ref{thm: main iter rate}
are satisfied on the event $\Mcal$. It follows by Theorem \ref{thm: main iter rate}
that on the event $\Mcal$, for any $K\geq1$, 
\[
\|\hbeta_{(K)}-\beta_{*}\|_{2}\leq2^{-K}c_{1}+2n^{-1/2}c_{3}^{-1/2}c_{5}\leq2^{-K}\kappa_{2}+2n^{-1/2}c_{3}^{-1/2}c_{5}.
\]

Notice that $1/(2\log2)<2$. Thus, for any $K\geq2\log n$, we have
that $2^{-K}\leq2^{-2\log n}\leq2^{-(\log n)/(2\log2)}=n^{-1/2}$.
Hence, on the event $\Mcal$, for any $K\geq2\log n$, we have 
\[
\|\hbeta_{(K)}-\beta_{*}\|_{2}\leq n^{-1/2}\kappa_{2}+4n^{-1/2}\kappa_{3}^{-1/2}c_{5}.
\]

Thus, $P(\sup_{K\geq2\log n}\|\hbeta_{(K)}-\beta_{*}\|_{2}\leq n^{-1/2}\kappa_{2}+4n^{-1/2}\kappa_{3}^{-1/2}c_{5})\geq P(\Mcal)$.
Since $\sup_{\|v-\beta_{*}\|_{2}\leq\kappa_{2}}\|H_{n}(v)\|_{2}=O_{P}(1)$,
the proof is complete.
\end{proof}
\begin{proof}[\textbf{Proof of Theorem \ref{thm: inference main}}]
We inherit all the notations and definitions from the proof of Corollary
\ref{cor: main rate}. Since $\varepsilon=\varepsilon_{n}=o(1)$,
we can assume $\varepsilon\in(0,\sqrt{\kappa_{3}}/8)$. We now fix
$K$, $\beta$ and $\Gamma$ such that $K\geq1+2\log n$, $\|\beta-\beta_{*}\|_{2}\leq A$
and $\|\Gamma-\Gamma_{*}\|\leq\varepsilon$. 

Let $K_{0}=K-1$. In the proof of Corollary \ref{cor: main rate},
we have showed that on the event $\Mcal$, 
\[
\|\Acal_{K_{0}}(\beta,\Gamma)-\beta_{*}\|_{2}\leq n^{-1/2}\kappa_{2}+4n^{-1/2}\kappa_{3}^{-1/2}\sup_{\|v-\beta_{*}\|_{2}\leq\kappa_{2}}\|H_{n}(v)\|_{2},
\]
where the event $\Mcal$ is defined by 
\[
\Mcal=\left\{ \sup_{\|v-\beta_{*}\|_{2}\leq\kappa_{2}}\|H_{n}(v)\|_{2}\leq\sqrt{n}\kappa_{3}/(32\kappa_{1})\right\} .
\]

Define $\eta_{0}=1-P(\Mcal)$. Fix an arbitrary $\eta\in(0,1)$, we
find a constant $M_{\eta}>0$ such that $P(\sup_{\|v-\beta_{*}\|_{2}\leq\kappa_{2}}\|H_{n}(v)\|_{2}>M_{\eta})\leq\eta$.
This is possible since we assume $\sup_{\|v-\beta_{*}\|_{2}\leq\kappa_{2}}\|H_{n}(v)\|_{2}=O_{P}(1)$.
Now we define the event $\bar{\Mcal}=\left\{ \sup_{\|v-\beta_{*}\|_{2}\leq\kappa_{2}}\|H_{n}(v)\|_{2}\leq\min\{M_{\eta},\ \sqrt{n}\kappa_{3}/(32\kappa_{1})\}\right\} $.
Notice that $P(\bar{\Mcal})\geq1-\max\{\eta_{0},\ \eta\}$ and $\bar{\Mcal}\subseteq\Mcal$.
Therefore, on the event $\bar{\Mcal}$, we have 
\[
\|\Acal_{K_{0}}(\beta,\Gamma)-\beta_{*}\|_{2}\leq n^{-1/2}C_{\eta},
\]
where $C_{\eta}=\kappa_{2}+4\kappa_{3}^{-1/2}\min\{M_{\eta},\ \sqrt{n}\kappa_{3}/(32\kappa_{1})\}$. 

Now we recall the basic decomposition (\ref{eq: key debias decomposition})
from the proof of Lemma \ref{lem: main iter}: 
\begin{align}
 & \Acal_{K}(\beta,\Gamma)-\beta_{*}\nonumber \\
 & =\Acal(\Acal_{K_{0}}(\beta,\Gamma),\Gamma)-\beta_{*}\nonumber \\
 & =(I-(\Gamma'\Gamma)^{-1}\Gamma'\Gamma_{*})(\Acal_{K_{0}}(\beta,\Gamma)-\beta_{*})\nonumber \\
 & \qquad-(\Gamma'\Gamma)^{-1}\Gamma'\left[n^{-1/2}H_{n}(\Acal_{K_{0}}(\beta,\Gamma))+\left(G(\Acal_{K_{0}}(\beta,\Gamma))-\Gamma_{*}(\Acal_{K_{0}}(\beta,\Gamma)-\beta_{*})\right)\right].\label{eq: inference thm eq 4}
\end{align}

By the proof of Theorem \ref{thm: main iter rate}, we have that $\|\Acal_{K_{0}}(\beta,\Gamma)-\beta_{*}\|_{2}\leq c_{1}$,
where $c_{1}=\min\{\sqrt{\kappa_{3}}/(8\kappa_{1}),\ \kappa_{2}\}$
(defined in the proof of Corollary \ref{cor: main rate}). Also in
the proof of Corollary \ref{cor: main rate}, we have that on the
event $\Mcal$, $\lambda_{\min}(\Gamma'\Gamma)\geq\kappa_{3}(7/8)^{2}$. 

Since $I-(\Gamma'\Gamma)^{-1}\Gamma'\Gamma_{*}=(\Gamma'\Gamma)^{-1}\Gamma'(\Gamma-\Gamma_{*})$
and $\|(\Gamma'\Gamma)^{-1}\Gamma'\|=1/\sqrt{\lambda_{\min}(\Gamma'\Gamma)}$,
we have that on the event $\bar{\Mcal}$, 
\begin{align*}
 & \|\Acal_{K}(\beta,\Gamma)-\beta_{*}+n^{-1/2}(\Gamma'\Gamma)^{-1}\Gamma'H_{n}(\beta_{*})\|_{2}\\
 & \leq\varepsilon\|(\Gamma'\Gamma)^{-1}\Gamma'\|\cdot\|\Acal_{K_{0}}(\beta,\Gamma)-\beta_{*}\|_{2}+\|(\Gamma'\Gamma)^{-1}\Gamma'\|\cdot n^{-1/2}\|H_{n}(\Acal_{K_{0}}(\beta,\Gamma))-H_{n}(\beta_{*})\|_{2}\\
 & \qquad+\kappa_{1}\|(\Gamma'\Gamma)^{-1}\Gamma'\|\cdot\|\Acal_{K_{0}}(\beta,\Gamma)-\beta_{*}\|_{2}^{2}\\
 & \leq\frac{8}{7}\kappa_{3}^{-1/2}\varepsilon\cdot\|\Acal_{K_{0}}(\beta,\Gamma)-\beta_{*}\|_{2}+\frac{8}{7}\kappa_{3}^{-1/2}\cdot n^{-1/2}\|H_{n}(\Acal_{K_{0}}(\beta,\Gamma))-H_{n}(\beta_{*})\|_{2}\\
 & \qquad+\frac{8}{7}\kappa_{3}^{-1/2}\kappa_{1}\cdot\|\Acal_{K_{0}}(\beta,\Gamma)-\beta_{*}\|_{2}^{2}\\
 & \leq\frac{8}{7}\kappa_{3}^{-1/2}\varepsilon n^{-1/2}C_{\eta}+\frac{8}{7}\kappa_{3}^{-1/2}\cdot n^{-1/2}\sup_{\|v\|_{2}\leq C_{\eta}}\|H_{n}(\beta_{*}+n^{-1/2}v)-H_{n}(\beta_{*})\|_{2}+\frac{8}{7}\kappa_{3}^{-1/2}\kappa_{1}n^{-1}C_{\eta}^{2}.
\end{align*}

By the continuity of $\Gamma\mapsto(\Gamma'\Gamma)^{-1}\Gamma'$,
it is not hard to see that there exists a constant $D>0$ depending
only on $\kappa_{3}$ such that $\|(\Gamma'\Gamma)^{-1}\Gamma'-(\Gamma_{*}'\Gamma_{*})^{-1}\Gamma_{*}'\|\leq D\|\Gamma-\Gamma_{*}\|$
for small enough $\varepsilon$. Therefore, 
\[
\left\Vert \left[(\Gamma'\Gamma)^{-1}\Gamma'-(\Gamma_{*}'\Gamma_{*})^{-1}\Gamma_{*}'\right]H_{n}(\beta_{*})\right\Vert _{2}\leq D\varepsilon\|H_{n}(\beta_{*})\|_{2}.
\]

The above two displays imply that on the even $\bar{\Mcal}$, 
\begin{multline*}
\|\Acal_{K}(\beta,\Gamma)-\beta_{*}+n^{-1/2}(\Gamma_{*}'\Gamma_{*})^{-1}\Gamma_{*}'H_{n}(\beta_{*})\|_{2}\\
\leq Dn^{-1/2}\varepsilon\|H_{n}(\beta_{*})\|_{2}+\frac{8}{7}\kappa_{3}^{-1/2}\varepsilon n^{-1/2}C_{\eta}+\frac{8}{7}\kappa_{3}^{-1/2}\cdot n^{-1/2}\sup_{\|v\|_{2}\leq C_{\eta}}\|H_{n}(\beta_{*}+n^{-1/2}v)-H_{n}(\beta_{*})\|_{2}\\
+\frac{8}{7}\kappa_{3}^{-1/2}\kappa_{1}n^{-1}C_{\eta}^{2}.
\end{multline*}

Since the above argument holds for any sample path on $\bar{\Mcal}$
and for any $(\beta,\Gamma)$, we have on the even $\bar{\Mcal}$,
\begin{multline*}
\sup_{\|\beta-\beta_{*}\|_{2}\leq A,\ \|\Gamma-\Gamma_{*}\|\leq\varepsilon}\|\Acal_{K}(\beta,\Gamma)-\beta_{*}+n^{-1/2}(\Gamma_{*}'\Gamma_{*})^{-1}\Gamma_{*}'H_{n}(\beta_{*})\|_{2}\\
\leq Dn^{-1/2}\varepsilon\|H_{n}(\beta_{*})\|_{2}+D_{1}\varepsilon n^{-1/2}+\frac{8}{7}\kappa_{3}^{-1/2}\cdot n^{-1/2}\sup_{\|v\|_{2}\leq C_{\eta}}\|H_{n}(\beta_{*}+n^{-1/2}v)-H_{n}(\beta_{*})\|_{2}\\
+n^{-1}D_{2},
\end{multline*}
where $D_{1}=\frac{8}{7}\kappa_{3}^{-1/2}C_{\eta}$ and $D_{2}=\frac{8}{7}\kappa_{3}^{-1/2}\kappa_{1}C_{\eta}^{2}$. 

Since $\eta_{0}=o(1)$, it follows that 
\begin{align*}
 & P\Biggl(\sup_{\|\beta-\beta_{*}\|_{2}\leq A,\ \|\Gamma-\Gamma_{*}\|\leq\varepsilon}\|\Acal_{K}(\beta,\Gamma)-\beta_{*}+n^{-1/2}(\Gamma_{*}'\Gamma_{*})^{-1}\Gamma_{*}'H_{n}(\beta_{*})\|_{2}\\
 & \qquad>Dn^{-1/2}\varepsilon\|H_{n}(\beta_{*})\|_{2}+D_{1}\varepsilon n^{-1/2}+\frac{8}{7}\kappa_{3}^{-1/2}\cdot n^{-1/2}\sup_{\|v\|_{2}\leq C_{\eta}}\|H_{n}(\beta_{*}+n^{-1/2}v)-H_{n}(\beta_{*})\|_{2}+n^{-1}D_{2}\Biggr)\\
 & \leq P(\bar{\Mcal}^{c})\leq\max\{\eta_{0},\ \eta\}\leq o(1)+\eta.
\end{align*}

Since $\eta>0$ is arbitrary and $\sup_{\|v\|_{2}\leq C}\|H_{n}(\beta_{*}+n^{-1/2}v)-H_{n}(\beta_{*})\|_{2}=o_{P}(1)$
for any $C>0$, the desired result follows.
\end{proof}

\section{Proof of results in Section \ref{sec: Jacobian estimate}}
\begin{proof}[\textbf{Proof of Lemma \ref{lem: BS approx}}]
Notice that by Taylor's theorem, we have 
\begin{align*}
G_{n}^{*}(b_{*}) & =n^{-1/2}H_{n}^{*}(b_{*})+G_{n}(b_{*})\\
 & =n^{-1/2}H_{n}^{*}(b_{*})+n^{-1/2}H_{n}(b_{*})+G(b_{*})\\
 & =n^{-1/2}H_{n}^{*}(b_{*})+n^{-1/2}H_{n}(b_{*})+G(b_{0})+\Gamma_{0}(b_{*}-b_{0})+a_{n}^{*}(b_{*}-b_{0})^{2},
\end{align*}
for some random variable $a_{n}^{*}$ with $|a_{n}^{*}|\leq\sup_{\beta}|d^{2}G(\beta)/d\beta^{2}|$.
On the other hand, 
\[
G_{n}(b_{0})=n^{-1/2}H_{n}(b_{0})+G(b_{0}).
\]

Since $G_{n}^{*}(b_{*})=G_{n}(b_{0})$, we have that
\[
n^{-1/2}H_{n}^{*}(b_{*})+n^{-1/2}H_{n}(b_{*})+G(b_{0})+\Gamma_{0}(b_{*}-b_{0})+a_{n}^{*}(b_{*}-b_{0})^{2}=n^{-1/2}H_{n}(b_{0})+G(b_{0}).
\]

The desired result follows.
\end{proof}
\begin{proof}[\textbf{Proof of Theorem \ref{thm: BS Jacobian}}]
Let $\delta=b_{*}-b_{0}$. By Lemma \ref{lem: BS approx}, we have
\begin{equation}
\hGamma_{0}-\Gamma_{0}=\frac{E\left(-n^{-1/2}H_{n}^{*}(b_{*})\delta\mid\Fcal\right)}{E(\delta^{2}\mid\Fcal)}-\Gamma_{0}=\frac{E\left((\Gamma_{0}\delta+\varepsilon)\delta\mid\Fcal\right)}{E(\delta^{2}\mid\Fcal)}-\Gamma_{0}=\frac{E\left(\delta\varepsilon\mid\Fcal\right)}{E(\delta^{2}\mid\Fcal)},\label{eq: thm Jacobian 4}
\end{equation}
where $\varepsilon=a_{n}^{*}\delta^{2}+n^{-1/2}B_{n}$ with $B_{n}=H_{n}(b_{*})-H_{n}(b_{0})$.
Fix an arbitrary number $\eta\in(0,1)$.

By the assumption, there exists a constant $C>0$ such that with probability
at least $1-\eta$, (1) $E\left(\sup_{\beta}|H_{n}^{*}(\beta)|\mid\Fcal\right)$
and $\sup_{\beta}|H_{n}(\beta)|$ are bounded by $C$ and (2) $\sup_{|x-y|\leq n^{-1/2}t}|H_{n}(x)-H_{n}(y)|\leq\eta$.
We define the following event
\begin{multline*}
\Mcal=\left\{ \sup_{|x-y|\leq n^{-1/2}\eta^{-1}}|H_{n}(x)-H_{n}(y)|\leq\eta\right\} \bigcap\left\{ E\left(\sup_{\beta}|H_{n}^{*}(\beta)|^{4}\mid\Fcal\right)\leq C\right\} \\
\bigcap\left\{ \sup_{\beta}|H_{n}(\beta)|\leq C\right\} \bigcap\left\{ E\left(\sup_{|x-y|\leq n^{-1/2}\eta^{-1}}|H_{n}^{*}(x)-H_{n}^{*}(y)|^{4}\mid\Fcal\right)\leq C\right\} \\
\bigcap\left\{ \rho_{2}<|\Gamma_{0}|<\rho_{3}\right\} \bigcap\left\{ E\left((H_{n}^{*}(b_{0}))^{2}\mid\Fcal\right)\geq\rho_{4}\right\} .
\end{multline*}

By assumption, $P(\Mcal)=1-\eta-o(1)$. Now we proceed in three steps,
in which we first characterize $\delta$ and then provide bounds for
$E(\delta\varepsilon\mid\Fcal)$ and $E(\delta^{2}\mid\Fcal)$. 

\textbf{Step 1:} provide preliminary bounds for $\delta$.

We first characterize the size of $\delta$. By Lemma \ref{lem: BS approx},
$-n^{-1/2}H_{n}^{*}(b_{*})=\Gamma_{0}\delta+a_{n}^{*}\delta^{2}+n^{-1/2}B_{n}$.
We rearrange the terms and obtain 
\[
\frac{a_{n}^{*}}{\Gamma_{0}}\delta^{2}+\delta+\frac{n^{-1/2}(H_{n}^{*}(b_{*})+B_{n})}{\Gamma_{0}}=0.
\]

By the quadratic formula, we have $2a_{n}^{*}\Gamma_{0}^{-1}\delta\in\{W_{1},W_{2}\}$,
where $W_{1}=\sqrt{1-4Q_{n}}-1$, $W_{2}=-\sqrt{1-4Q_{n}}-1$ and
$Q_{n}=n^{-1/2}a_{n}^{*}(H_{n}^{*}(b_{*})+B_{n})\Gamma_{0}^{-2}$.
Notice that $1\leq|W_{2}|\leq2$ and 
\begin{equation}
|W_{1}|=\frac{\left|\sqrt{1-4Q_{n}}-1\right|}{1}=\frac{\left|-4Q_{n}\right|}{\sqrt{1-4Q_{n}}+1}\leq4|Q_{n}|.\label{eq: thm Jacobian 7}
\end{equation}

On the event $\Mcal$, we have 
\begin{align*}
E\left(|Q_{n}|^{3}\mid\Fcal\right) & \leq\left(n^{-1/2}\rho_{1}\rho_{2}^{-2}\right)^{3}E\left([|H_{n}^{*}(b_{*})|+2C]^{3}\mid\Fcal\right)\\
 & \overset{\text{(i)}}{\leq}4\left(n^{-1/2}\rho_{1}\rho_{2}^{-2}\right)^{3}E\left(|H_{n}^{*}(b_{*})|^{3}+8C^{3}\mid\Fcal\right)\\
 & \leq4\left(n^{-1/2}\rho_{1}\rho_{2}^{-2}\right)^{3}\left(C^{3}+8C^{3}\right)=n^{-3/2}\kappa
\end{align*}
with $\kappa=36\rho_{1}^{3}\rho_{2}^{-6}$, where (i) follows by the
elementary inequality $(a+b)^{3}\leq4(a^{3}+b^{3})$. Since $b_{*}$
is chosen to be the closest solution to $b_{0}$, it follows that
$2a_{n}^{*}\Gamma_{0}^{-1}\delta=W_{2}$ only if $|W_{2}|\leq|W_{1}|$.
Due to $|W_{2}|\geq1$ and (\ref{eq: thm Jacobian 7}), $|W_{2}|\leq|W_{1}|$
implies $|Q_{n}|\geq1/4$. Therefore, 

\begin{multline}
P\left(2a_{n}^{*}\Gamma_{0}^{-1}\delta\neq W_{1}\mid\Fcal\right)\leq P\left(|Q_{n}|\geq1/4\mid\Fcal\right)=P\left(|Q_{n}|^{3}\geq1/64\mid\Fcal\right)\\
\leq64E\left(|Q_{n}|^{3}\mid\Fcal\right)\leq64\kappa n^{-3/2}.\label{eq: thm Jacobian 9}
\end{multline}

By (\ref{eq: thm Jacobian 7}) and the definition of $Q_{n}$, we
have that 
\begin{align}
|\delta|\oneb\{2a_{n}^{*}\Gamma_{0}^{-1}\delta=W_{1}\} & =\left|\frac{W_{1}}{2a_{n}^{*}\Gamma_{0}^{-1}}\right|\oneb\{2a_{n}^{*}\Gamma_{0}^{-1}\delta=W_{1}\}\nonumber \\
 & \leq\left|\frac{2Q_{n}}{a_{n}^{*}\Gamma_{0}^{-1}}\right|\oneb\{2a_{n}^{*}\Gamma_{0}^{-1}\delta=W_{1}\}\nonumber \\
 & \leq2n^{-1/2}\rho_{3}\left|H_{n}^{*}(b_{*})+B_{n}\right|\oneb\{2a_{n}^{*}\Gamma_{0}^{-1}\delta=W_{1}\}.\label{eq: thm Jacobian 10}
\end{align}

By construction, $|\delta|\leq\bC$. Thus, for any $r\in[1,4]$, we
observe that 
\begin{align*}
 & E(|\delta|^{r}\mid\Fcal)\\
 & =E(|\delta|^{r}\oneb\{2a_{n}^{*}\Gamma_{0}^{-1}\delta=W_{1}\}\mid\Fcal)+E(|\delta|^{r}\oneb\{2a_{n}^{*}\Gamma_{0}^{-1}\delta\neq W_{1}\}\mid\Fcal)\\
 & \leq2^{r}n^{-r/2}\rho_{3}^{r}E\left(\left(|H_{n}^{*}(b_{*})|+2C\right)^{r}\times\oneb\{2a_{n}^{*}\Gamma_{0}^{-1}\delta=W_{1}\}\mid\Fcal\right)+\bC^{r}P\left(2a_{n}^{*}\Gamma_{0}^{-1}\delta\neq W_{1}\mid\Fcal\right)\\
 & \overset{\text{(i)}}{\leq}2^{2r-1}n^{-r/2}\rho_{3}^{r}E\left(\left(|H_{n}^{*}(b_{*})|^{r}+2^{r}C^{r}\right)\times\oneb\{2a_{n}^{*}\Gamma_{0}^{-1}\delta=W_{1}\}\mid\Fcal\right)+\bC^{r}P\left(2a_{n}^{*}\Gamma_{0}^{-1}\delta\neq W_{1}\mid\Fcal\right)\\
 & \overset{\text{(ii)}}{\leq}2^{2r-1}n^{-r/2}\rho_{3}^{r}C^{r}(1+2^{r})+\bC^{r}\times64\kappa n^{-3/2},
\end{align*}
where (i) follows by the elementary inequality $(a+b)^{r}\leq2^{r-1}(a^{r}+b^{r})$
for $a,b\geq0$ and (ii) follows by the definition of $\Mcal$ and
(\ref{eq: thm Jacobian 9}). Therefore, there exists a constant $C_{0}>0$
such that
\begin{equation}
E(|\delta|^{r}\mid\Fcal)\leq C_{0}\left(n^{-r/2}+n^{-3/2}\right)\qquad\forall r\in[1,4].\label{eq: thm Jacobian 11}
\end{equation}

\textbf{Step 2:} derive a lower bound for $E(\delta^{2}\mid\Fcal)$.

By (\ref{eq: thm Jacobian 11}), we have 
\begin{equation}
P\left(|\delta|>n^{-1/2}\eta^{-1}\mid\Fcal\right)\leq n^{1/2}\eta E(|\delta|\mid\Fcal)\leq C_{0}\left(1+n^{-1}\right)\eta\leq2C_{0}\eta.\label{eq: thm Jacobian 12}
\end{equation}

Clearly, $|B_{n}|\leq2C$ on the event $\Mcal$. Thus, 
\begin{align}
E\left(B_{n}^{2}\mid\Fcal\right) & =E\left(B_{n}^{2}\oneb\{|\delta|>n^{-1/2}\eta^{-1}\}\mid\Fcal\right)+E\left(B_{n}^{2}\oneb\{|\delta|\leq n^{-1/2}\eta^{-1}\}\mid\Fcal\right)\nonumber \\
 & \leq4C^{2}P\left(|\delta|>n^{-1/2}\eta^{-1}\mid\Fcal\right)+E\left(B_{n}^{2}\oneb\{|\delta|\leq n^{-1/2}\eta^{-1}\}\mid\Fcal\right)\nonumber \\
 & \overset{\text{(i)}}{\leq}4C^{2}P\left(|\delta|>n^{-1/2}\eta^{-1}\mid\Fcal\right)+\eta^{2}E\left(\oneb\{|\delta|\leq n^{-1/2}\eta^{-1}\}\mid\Fcal\right)\nonumber \\
 & \leq8C^{2}C_{0}\eta+\eta^{2},\label{eq: thm Jacobian 13}
\end{align}
where (i) follows by the fact that on the event $\Mcal\bigcap\{|\delta|\leq n^{-1/2}\eta^{-1}\}$,
$|B_{n}|\leq\sup_{|x-y|\leq n^{-1/2}\eta^{-1}}|H_{n}(x)-H_{n}(y)|\leq\eta$.
Similarly, we have 
\begin{align}
 & E\left((H_{n}^{*}(b_{*})-H_{n}^{*}(b_{0}))^{2}\mid\Fcal\right)\nonumber \\
 & =E\left((H_{n}^{*}(b_{*})-H_{n}^{*}(b_{0}))^{2}\oneb\{|\delta|>n^{-1/2}\eta^{-1}\}\mid\Fcal\right)\nonumber \\
 & \qquad\qquad+E\left((H_{n}^{*}(b_{*})-H_{n}^{*}(b_{0}))^{2}\oneb\{|\delta|\leq n^{-1/2}\eta^{-1}\}\mid\Fcal\right)\nonumber \\
 & \overset{\text{(i)}}{\leq}\sqrt{E\left((H_{n}^{*}(b_{*})-H_{n}^{*}(b_{0}))^{4}\mid\Fcal\right)\times P\left(|\delta|>n^{-1/2}\eta^{-1}\mid\Fcal\right)}\nonumber \\
 & \qquad\qquad+E\left((H_{n}^{*}(b_{*})-H_{n}^{*}(b_{0}))^{2}\oneb\{|\delta|\leq n^{-1/2}\eta^{-1}\}\mid\Fcal\right)\nonumber \\
 & \overset{\text{(ii)}}{\leq}\sqrt{8E\left(\left[|H_{n}^{*}(b_{*})|^{4}+|H_{n}^{*}(b_{0})|^{4}\right]\mid\Fcal\right)\times P\left(|\delta|>n^{-1/2}\eta^{-1}\mid\Fcal\right)}\nonumber \\
 & \qquad\qquad+E\left((H_{n}^{*}(b_{*})-H_{n}^{*}(b_{0}))^{2}\oneb\{|\delta|\leq n^{-1/2}\eta^{-1}\}\mid\Fcal\right)\nonumber \\
 & \overset{\text{(iii)}}{\leq}\sqrt{16C\times2C_{0}\eta}+E\left((H_{n}^{*}(b_{*})-H_{n}^{*}(b_{0}))^{2}\oneb\{|\delta|\leq n^{-1/2}\eta^{-1}\}\mid\Fcal\right)\nonumber \\
 & \overset{\text{(iv)}}{\leq}\sqrt{32CC_{0}\eta}+E\left(\eta^{2}\oneb\{|\delta|\leq n^{-1/2}\eta^{-1}\}\mid\Fcal\right)\nonumber \\
 & \leq\sqrt{32CC_{0}\eta}+\eta^{2},\label{eq: thm Jacobian 14}
\end{align}
where (i) follows by Cauchy-Schwarz inequality, (ii) follows by the
elementary inequality $(a-b)^{4}\leq(|a|+|b|)^{4}\leq8(|a|^{4}+|b|^{4})$
for $a,b\in\RR$, (iii) follows by (\ref{eq: thm Jacobian 12}) and
the definition of $\Mcal$ and finally (iv) follows by the fact that
on the event $\Mcal\bigcap\{|\delta|\leq n^{-1/2}\eta^{-1}\}$, 
\[
\left|H_{n}^{*}(b_{*})-H_{n}^{*}(b_{0})\right|\leq\sup_{|x-y|\leq n^{-1/2}\eta^{-1}}\left|H_{n}^{*}(x)-H_{n}^{*}(y)\right|\leq\eta.
\]

By (\ref{eq: thm Jacobian 13}) and (\ref{eq: thm Jacobian 14}),
there exists a constant $C_{1}>0$ independent of $\eta$ such that
\begin{equation}
\max\left\{ E\left((H_{n}^{*}(b_{*})-H_{n}^{*}(b_{0}))^{2}\mid\Fcal\right),\ E\left(B_{n}^{2}\mid\Fcal\right)\right\} \leq C_{1}^{2}\sqrt{\eta}.\label{eq: thm Jacobian 14.5}
\end{equation}
We notice that
\begin{align}
E(\delta^{2}\mid\Fcal) & \geq E(\delta^{2}\oneb\{2a_{n}^{*}\Gamma_{0}^{-1}\delta=W_{1}\}\mid\Fcal)\nonumber \\
 & =E\left(\frac{W_{1}^{2}}{4(a_{n}^{*})^{2}\Gamma_{0}^{-2}}\times\oneb\{2a_{n}^{*}\Gamma_{0}^{-1}\delta=W_{1}\}\mid\Fcal\right)\nonumber \\
 & \geq E\left(\frac{4Q_{n}^{2}}{4(a_{n}^{*})^{2}\Gamma_{0}^{-2}}\times\oneb\{2a_{n}^{*}\Gamma_{0}^{-1}\delta=W_{1}\}\mid\Fcal\right)\nonumber \\
 & =\Gamma_{0}^{2}n^{-1}E\left((H_{n}^{*}(b_{*})+B_{n})^{2}\times\oneb\{2a_{n}^{*}\Gamma_{0}^{-1}\delta=W_{1}\}\mid\Fcal\right)\nonumber \\
 & \geq\rho_{2}^{2}n^{-1}E\left((H_{n}^{*}(b_{*})+B_{n})^{2}\times\oneb\{2a_{n}^{*}\Gamma_{0}^{-1}\delta=W_{1}\}\mid\Fcal\right).\label{eq: thm Jacobian 15}
\end{align}

For a random variable $R$, we can verify that $\|R\|:=\sqrt{E(R^{2}\times\oneb\{2a_{n}^{*}\Gamma_{0}^{-1}\delta=W_{1}\}\mid\Fcal)}$
is a semi-norm and we hence can apply the triangular inequality $\|R^{(1)}+R^{(2)}+R^{(3)}\|\geq\|R^{(1)}\|-\|R^{(2)}\|-\|R^{(3)}\|$
for any norm. Therefore, we notice that 
\begin{align*}
 & \sqrt{E\left((H_{n}^{*}(b_{*})+B_{n})^{2}\times\oneb\{2a_{n}^{*}\Gamma_{0}^{-1}\delta=W_{1}\}\mid\Fcal\right)}\\
 & =\sqrt{E\left(\left[H_{n}^{*}(b_{0})+(H_{n}^{*}(b_{*})-H_{n}^{*}(b_{0}))+B_{n}\right]^{2}\times\oneb\{2a_{n}^{*}\Gamma_{0}^{-1}\delta=W_{1}\}\mid\Fcal\right)}\\
 & \geq\sqrt{E\left((H_{n}^{*}(b_{0}))^{2}\times\oneb\{2a_{n}^{*}\Gamma_{0}^{-1}\delta=W_{1}\}\mid\Fcal\right)}\\
 & \qquad-\sqrt{E\left((H_{n}^{*}(b_{*})-H_{n}^{*}(b_{0}))^{2}\times\oneb\{2a_{n}^{*}\Gamma_{0}^{-1}\delta=W_{1}\}\mid\Fcal\right)}\\
 & \qquad-\sqrt{E\left(B_{n}^{2}\times\oneb\{2a_{n}^{*}\Gamma_{0}^{-1}\delta=W_{1}\}\mid\Fcal\right)}\\
 & \geq\sqrt{E\left((H_{n}^{*}(b_{0}))^{2}\times\oneb\{2a_{n}^{*}\Gamma_{0}^{-1}\delta=W_{1}\}\mid\Fcal\right)}-\sqrt{E\left((H_{n}^{*}(b_{*})-H_{n}^{*}(b_{0}))^{2}\mid\Fcal\right)}-\sqrt{E\left(B_{n}^{2}\mid\Fcal\right)}\\
 & \overset{\text{(i)}}{\geq}\sqrt{E\left((H_{n}^{*}(b_{0}))^{2}\times\oneb\{2a_{n}^{*}\Gamma_{0}^{-1}\delta=W_{1}\}\mid\Fcal\right)}-2C_{1}\eta^{1/4},
\end{align*}
where (i) follows by (\ref{eq: thm Jacobian 14.5}). We now observe
that 
\begin{align*}
 & E\left((H_{n}^{*}(b_{0}))^{2}\times\oneb\{2a_{n}^{*}\Gamma_{0}^{-1}\delta=W_{1}\}\mid\Fcal\right)\\
 & =E\left((H_{n}^{*}(b_{0}))^{2}\mid\Fcal\right)-E\left((H_{n}^{*}(b_{0}))^{2}\times\oneb\{2a_{n}^{*}\Gamma_{0}^{-1}\delta\neq W_{1}\}\mid\Fcal\right)\\
 & \overset{\text{(i)}}{\geq}\rho_{4}-\sqrt{E\left((H_{n}^{*}(b_{0}))^{4}\mid\Fcal\right)\times P\left(2a_{n}^{*}\Gamma_{0}^{-1}\delta\neq W_{1}\mid\Fcal\right)}\\
 & \overset{\text{(ii)}}{\geq}\rho_{4}-\sqrt{C\times64\kappa n^{-3/2}},
\end{align*}
where (i) follows by Cauchy-Schwarz inequality and the definition
of $\Mcal$ and (ii) follows by (\ref{eq: thm Jacobian 9}). Hence,
for $n\geq C_{2}$ with $C_{2}=(256C\kappa\rho_{4}^{-2})^{2/3}$,
the above two displays imply 
\[
\sqrt{E\left((H_{n}^{*}(b_{*})+B_{n})^{2}\times\oneb\{2a_{n}^{*}\Gamma_{0}^{-1}\delta=W_{1}\}\mid\Fcal\right)}\geq\sqrt{\rho_{4}/2}-2C_{1}\eta^{1/4}.
\]

By (\ref{eq: thm Jacobian 15}), this means that for $n\geq C_{2}$,
\begin{equation}
E(\delta^{2}\mid\Fcal)\geq\rho_{2}^{2}n^{-1}\left(\max\left\{ \sqrt{\rho_{4}/2}-2C_{1}\eta^{1/4},0\right\} \right)^{2}.\label{eq: thm Jacobian 16}
\end{equation}

\textbf{Step 3:} derive an upper bound for $E(\varepsilon\delta\mid\Fcal)$. 

Notice that 
\begin{align}
 & \left|E(\varepsilon\delta\mid\Fcal)\right|\nonumber \\
 & =\left|E\left(a_{n}^{*}\delta^{3}+n^{-1/2}B_{n}\delta\mid\Fcal\right)\right|\nonumber \\
 & \leq\rho_{1}E(|\delta|^{3}\mid\Fcal)+n^{-1/2}E(|B_{n}\delta|\mid\Fcal)\nonumber \\
 & \leq\rho_{1}E(|\delta|^{3}\mid\Fcal)+n^{-1/2}\sqrt{E(B_{n}^{2}\mid\Fcal)\times E(\delta^{2}\mid\Fcal)}\nonumber \\
 & \overset{\text{(i)}}{\leq}\rho_{1}E(|\delta|^{3}\mid\Fcal)+n^{-1/2}\sqrt{C_{1}^{2}\sqrt{\eta}\times E(\delta^{2}\mid\Fcal)}\nonumber \\
 & \overset{\text{(i)}}{\leq}2\rho_{1}C_{0}n^{-3/2}+n^{-1/2}\sqrt{C_{1}^{2}\eta\times C_{0}\left(n^{-1}+n^{-3/2}\right)},\label{eq: thm Jacobian 17}
\end{align}
where (i) follows by (\ref{eq: thm Jacobian 14.5}) and (ii) follows
by (\ref{eq: thm Jacobian 11}). Finally, we combine (\ref{eq: thm Jacobian 16})
and (\ref{eq: thm Jacobian 17}), obtaining that on the event $\Mcal$,
for $n\geq C_{2}$, 
\[
\left|\frac{E(\varepsilon\delta\mid\Fcal)}{E(\delta^{2}\mid\Fcal)}\right|\leq\frac{2\rho_{1}C_{0}n^{-1/2}+\sqrt{C_{1}^{2}\eta\times C_{0}\left(1+n^{-1/2}\right)}}{\rho_{2}^{2}\left(\max\left\{ \sqrt{\rho_{4}/2}-2C_{1}\eta^{1/4},0\right\} \right)^{2}}.
\]

Therefore, in light of (\ref{eq: thm Jacobian 4}), for $n\geq C_{2}$,
\[
P\left(|\hGamma_{0}-\Gamma_{0}|>\frac{2\rho_{1}C_{0}n^{-1/2}+\sqrt{C_{1}^{2}\eta\times C_{0}\left(1+n^{-1/2}\right)}}{\rho_{2}^{2}\left(\max\left\{ \sqrt{\rho_{4}/2}-2C_{1}\eta^{1/4},0\right\} \right)^{2}}\right)\leq P(\Mcal^{c})=\eta+o(1).
\]

Since $\eta\in(0,1)$ is arbitrary, the desired result follows.
\end{proof}

\section{Proof of results in Section \ref{sec: IVQR}}
\begin{proof}[\textbf{Proof of Theorem \ref{thm: rate of convergence}}]
Fix an arbitrary $\eta>0$. Since $G_{n}(\beta)=G(\beta)+n^{-1/2}H_{n}(\beta)$,
we have that $\|G(\hat{\beta})+n^{-1/2}H_{n}(\hat{\beta})\|_{\infty}=o_{P}(1)$.
Thus, 
\[
\|G(\hbeta)\|_{\infty}\leq o_{P}(1)+n^{-1/2}\|H_{n}(\hbeta)\|_{\infty}\leq o_{P}(1)+\sup_{\beta\in\Bcal}\|n^{-1/2}H_{n}(\beta)\|_{\infty}=o_{P}(1).
\]

By Assumption \ref{assu: ID}, we have that $\|\hat{\beta}-\beta\|_{2}\leq\eta$
with probability approaching one; otherwise, we would have that $\|G(\hat{\beta})\|_{\infty}>C_{\eta}$
with non-vanishing probability, contradicting $\|G(\hat{\beta})\|_{\infty}=o_{P}(1)$.
Since $\eta>0$ is arbitrary, we have $\hat{\beta}=\beta_{*}+o_{P}(1)$.

Define the event $\Mcal=\{\|\hbeta-\beta_{*}\|_{2}\leq\min\{c_{1},c\}\}$.
By Assumption \ref{assu: Donsker} and $\hbeta=\beta_{*}+o_{P}(1)$,
we have $P(\Mcal)\rightarrow1$. Now we have that on the event $\Mcal$,
\[
\|G(\hbeta)\|_{\infty}\leq\|G_{n}(\hbeta)\|_{\infty}+n^{-1/2}\|H_{n}(\hbeta)\|_{\infty}\leq\|G_{n}(\hbeta)\|_{\infty}+n^{-1/2}\sup_{\|\beta-\beta_{*}\|_{2}\leq c}\|H_{n}(\beta)\|_{\infty}
\]
and 
\[
\|G(\hbeta)\|_{\infty}\geq\frac{\|G(\hbeta)\|_{2}}{\sqrt{L}}\geq\frac{c_{2}\|\hbeta-\beta_{*}\|_{2}}{\sqrt{L}}.
\]

Therefore, on the event $\Mcal$, we have that 
\[
\|\hbeta-\beta_{*}\|_{2}\leq c_{2}^{-1}\sqrt{L}\left(\|G_{n}(\hbeta)\|_{\infty}+n^{-1/2}\sup_{\|\beta-\beta_{*}\|_{2}\leq c}\|H_{n}(\beta)\|_{\infty}\right).
\]

The desired result follows by $P(\Mcal)\rightarrow1$. 
\end{proof}

\begin{proof}[\textbf{Proof of Lemma \ref{lem: bnd self normaliz}}]
Recall $\varepsilon_{i}=Y_{i}-X_{i}'\beta_{*}$ and $EZ_{i}(\mathbf{1}\{\varepsilon_{i}\leq0\}-\tau)=0$.
We fix $j\in\{1,...,L\}$. Let $L_{n}=C_{1}n$ and $B_{n}^{2}=C_{2}n$,
where $C_{1}=E|Z_{1,j}|^{3}\left|\mathbf{1}\{\varepsilon_{i}\leq0\}-\tau\right|^{3}$
and $C_{2}=EZ_{1,j}^{2}\left(\mathbf{1}\{\varepsilon_{i}\leq0\}-\tau\right)^{2}$.
We apply Theorem 7.4 of \citet{pena2008self} with $\delta=1$. Hence,
for any $0\leq x\leq C_{2}C_{1}^{-1/3}n^{1/6}$, 
\[
P\left(\frac{\left|\sum_{i=1}^{n}Z_{i,j}\left(\mathbf{1}\{\varepsilon_{i}\leq0\}-\tau\right)\right|}{\sqrt{\sum_{i=1}^{n}Z_{i,j}^{2}\left(\mathbf{1}\{\varepsilon_{i}\leq0\}-\tau\right)^{2}}}>x\right)\leq2\left(1+A\left(\frac{1+x}{C_{2}C_{1}^{-1/3}n^{1/6}}\right)^{3}\right)\left(1-\Phi(x)\right),
\]
where $A$ is an absolute constant. By the union bound, it follows
that for any $0\leq x\leq C_{2}C_{1}^{-1/3}n^{1/6}$, 
\[
P\left(\max_{1\leq j\leq L}\frac{\left|\sum_{i=1}^{n}Z_{i,j}\left(\mathbf{1}\{\varepsilon_{i}\leq0\}-\tau\right)\right|}{\sqrt{\sum_{i=1}^{n}Z_{i,j}^{2}\left(\mathbf{1}\{\varepsilon_{i}\leq0\}-\tau\right)^{2}}}>x\right)\leq2L\left(1+A\left(\frac{1+x}{C_{2}C_{1}^{-1/3}n^{1/6}}\right)^{3}\right)\left(1-\Phi(x)\right).
\]

Now we take $x=\Phi^{-1}(1-\alpha/n)$ for $\alpha\geq1/n$. Then
clearly, $x\leq\Phi^{-1}(1-n^{-2})\asymp\sqrt{\log n}\ll n^{1/6}$.
Therefore, for large $n$ (satisfying $\Phi^{-1}(1-n^{-2})\leq C_{2}C_{1}^{-1/3}n^{1/6})$,
\[
P\left(\max_{1\leq j\leq L}\frac{\left|\sum_{i=1}^{n}Z_{i,j}\left(\mathbf{1}\{\varepsilon_{i}\leq0\}-\tau\right)\right|}{\sqrt{\sum_{i=1}^{n}Z_{i,j}^{2}\left(\mathbf{1}\{\varepsilon_{i}\leq0\}-\tau\right)^{2}}}>\Phi^{-1}(1-\alpha/n)\right)\leq2L\alpha n^{-1}\left(1+a_{n}\right),
\]
where $a_{n}=o(1)$ only depends on $C_{1}$ and $C_{2}$. Since $\mathbf{1}\{\varepsilon_{i}\leq0\}-\tau\in\{-\tau,1-\tau\}$,
it follows that $(\mathbf{1}\{\varepsilon_{i}\leq0\}-\tau)^{2}\leq\max\{\tau^{2},(1-\tau)^{2}\}\leq1$
and thus
\[
\max_{1\leq j\leq L}\sum_{i=1}^{n}Z_{i,j}^{2}\left(\mathbf{1}\{\varepsilon_{i}\leq0\}-\tau\right)^{2}\leq\max_{1\leq j\leq L}\sum_{i=1}^{n}Z_{i,j}^{2}.
\]

Therefore, we have that 
\[
P\left(\|G_{n}(\beta_{*})\|_{\infty}>\Phi^{-1}(1-\alpha/n)n^{-1}\sqrt{\max_{1\leq j\leq L}\sum_{i=1}^{n}Z_{i,j}^{2}}\right)\leq2L\alpha n^{-1}\left(1+a_{n}\right).
\]

The proof is complete.
\end{proof}

\end{document}